\documentclass[iop]{emulateapj}
\usepackage{natbib,graphicx,amsmath,longtable,amssymb}

\def\nenc{N_{\rm enc}}

\def\ml{M_{\rm sub}}
\def\nl{n_{\rm subhalo}}

\def\rs{r_{\rm s}}

\def\msun{M_{\odot}}
\shorttitle{Detecting Subhalos with With Streams }

\shortauthors{Joo Heon Yoon}
\slugcomment{ApJ Accepted}

\begin{document}

\title{
Clumpy Streams from Clumpy Halos: \\
Detecting Missing Satellites with Cold Stellar Structures}

\author{Joo Heon Yoon\altaffilmark{1}$^\star$}
\author{Kathryn V. Johnston\altaffilmark{1}}
\author{David W. Hogg\altaffilmark{2}}

\altaffiltext{1}{Department of Astronomy, Columbia University, New York 10027, USA}
\altaffiltext{2}{Center for Cosmology and Particle Physics, Department of Physics, New York University, New York 10003, USA}
\email{$^\star$jhyoon@astro.columbia.edu}

\begin{abstract}
  Dynamically cold stellar streams are ideal probes of the
  gravitational field of the Milky Way.  This paper re-examines the
  question of how such streams might be used to test for the presence
  of ``missing satellites'' --- the many thousands of dark-matter
  subhalos with masses $10^5-10^7 \rm M_\odot$ which are seen to orbit
  within Galactic-scale dark-matter halos in simulations of structure
  formation in $\Lambda$CDM cosmologies.  Analytical estimates of the
  frequency and energy scales of stream encounters 
  indicate
  that these missing satellites should have a negligible effect on hot
  debris structures, such as the tails from the Sagittarius dwarf
  galaxy.  However, long cold streams, such as the structure known as GD-1 or those from the globular
  cluster Palomar 5 (Pal 5)  
  are expected to suffer many tens of direct
  impacts from missing satellites during their lifetimes.  Numerical
  experiments confirm that these impacts create gaps in the debris'
  orbital energy distribution, which will evolve into degree- and
  sub-degree- scale fluctuations in surface density over the age of
  the debris.  Maps of Pal 5's own stream contain surface density
  fluctuations on these scales. The presence and frequency of these inhomogeneities
  suggests the existence of a population of missing satellites 
  in numbers predicted in the standard $\Lambda$CDM cosmologies.

\end{abstract}
 
\clearpage

\section{Introduction}

Standard $\Lambda$CDM models of the Universe allow us to explain
structure formation on large scales. However, they predict an order of
magnitude more dark-matter subhalos within the halos of typical
galaxies than the number of known satellite galaxies orbiting the
Milky Way \citep{Klypin1999a,Moore1999a,Diemand2007a,Springel2008a}.  Recent,
large-area stellar surveys have discovered dozens of new satellite
galaxies, most notably using the Sloan Digital Sky Survey \citep[SDSS,
e.g.][]{Willman2005a,Belokurov2006a,Belokurov2007a,
  Zucker2006a,Irwin2007a,Koposov2007a,Walsh2007a} but the number discrepancy
between simulated dark-matter subhalos and observed satellite
populations is still significant.  This discrepancy can partially be
explained by accounting for the the incomplete sky-coverage of SDSS
and the distance-dependent limit on this survey's sensitivity to
low-surface brightness objects \citep{Koposov2008a,Tollerud2008a}.  
Indeed, models which take this into account and 
consider diffuse, (i.e. undetectable) satellite galaxies can reconcile the number
counts for subhalos \citep{Bullock2010b}. However, when they impose the
suppression of stellar populations in low mass subhalos (which have
masses below $5 \times10^8 \rm \msun$) 
the number of undetected galaxies
significantly declines and the
prediction of numerous purely dark-matter subhalos less massive than
$5 \times 10^8 \rm \msun$ remains. 

There could be a genuine absence
of ``missing satellites" in the inner halo due to destruction by disk shocks, as illustrated in the calculations \citet{DOnghia2010a}.
However, note that these analytic descriptions of disk shocking based on the energy criterion are known to overestimate disruption rates of subhalos significantly\citep{Goerdt2007a}. Once these destructive effects are accurately accounted for, proof of the existence (or lack) of these ``missing satellites'' 
could provide an important constraint on the nature of dark
matter, which sets the minimum scale for the formation of dark-matter
subhalos \citep[e.g.][]{Hooper2007a}.

Along with the discovery of new satellite galaxies, SDSS has also
uncovered a multitude of stellar structures in the Milky Way halo from
disrupting globular clusters or satellite galaxies. In many cases, the
debris is dynamically cold and distributed narrowly in space
\citep{Odenkirchen2001a,Belokurov2006c,Lauchner2006a,Grillmair2006a,Grillmair2006c, Grillmair2006d,Grillmair2009a}. Such cold stellar streams should be
sensitive probes of the gravitational potential.  On global scales,
they can be used to constrain the radial profile, shape and
orientation of the Milky Way's triaxial dark-matter halo
\citep[e.g.][]{Johnston1999a,Ibata2002a,Johnston2005a,Binney2008a,Eyre2010a,Koposov2010a,Law2010a}.
The presence of dark-matter subhalos would add asymmetries to the
global potential over a range of smaller scales which will perturb
these cold streams or even destroy them.  Hence, if the missing
satellites do exist they will add random uncertainties to any
stellar-dynamical assessment of the global potential.

Gravitational lensing has been suggested to be a useful tool to probe the presence of
subhalos
\citep{Chiba2002a,Metcalf2002a,Chen2003a,Moustakas2003a,Metcalf2004a,Keeton2009a,Riehm2009a,Xu2009a}. 
These investigations conclude that flux ratio anomalies in lensed images or 
lensing time delays could be caused 
by dark-matter subhalos,
though the constraints are limited by our knowledge of
the spatial distribution
of subhalos. However, this method is only applicable to the most massive and most centrally concentrated dark-matter halos, and not to galaxies like the Milky Way more generally.

The effect of dark-matter subhalos on stellar streams has been
explored in several previous studies.  \citet{Ibata2002a} showed that
debris from the destruction of a $10^6 \rm M_{\odot}$ globular cluster
should be affected by heating due to repeated close encounters of
subhalos and concluded that this effect could be detectable with
future astrometric surveys.
Moreover, \citet{Quinn2008a} found that the inhomogeneities seen
in Pal 5's tidal tails could not be accounted for
in simulations evolved in a smooth potential.
For the streams of larger satellites like
the Sagittarius dwarf galaxy (hereafter Sgr), \citet{Johnston2002a} found that
although stars in the debris are scattered by encounters with dark
matter subhalos, the thickness of the current Sgr stream could be
explained as being due to the Large Magellanic Cloud alone.
\citet{Siegal-Gaskins2008a} tested the additional influence of different host
potentials on debris from satellite galaxies and pointed out that
while subhalos can shift the positions of streams and cause clumpy
structures, the shape of the halo potential and orbital path can have
an overall comparable effect.  Most recently, \citet{Carlberg2009a}
modeled a simplified stream on a circular orbit and concluded that
dynamically old ($>$ 3Gyr) stellar streams cannot survive in the presence
of subhalos with the masses and numbers predicted by $\Lambda$CDM.

These previous works point to stellar streams as perhaps the most
powerful way to find the missing satellites. However, 
none of these investigations separated 
the effect on streams of the known (and therefore
uninteresting!) satellites with masses $> 10^8 \msun$ 
\citep[][and references therein]{Bovill2009a,Bullock2010a} from those that
are ``missing'' (the pure dark-matter subhalos).
In this study we construct a framework for understanding stream
inhomogeneities by first isolating and dissecting the characteristics
of disturbances caused by dark-matter subhalos alone.  In contrast to
previous work, which looked at the overall response of cold streams to
the complete $\Lambda$CDM subhalo mass spectrum, we look at the
expected frequency, influence and characteristic observable signatures
of subhalos in each mass decade separately.  We also
contrast the response of different streams to the same
masses, from ribbons such as Pal 5 to the giant stream from Sgr.
Our twin aims are: (i) to understand
with which streams we are most likely to be able to conclusively prove
the existence or absence of missing satellites: and (ii) to learn how 
signatures of missing satellites that are apparent in streams might be interpreted.

It should be noted that the
discovery of very cold streams from globular clusters has inspired
discussions of how the intrinsic properties of stellar streams themselves
could cause inhomogeneities in their density distributions 
\citep{Kupper2008a,Kupper2010a,Quillen2010a} and these self-induced 
fluctuations could confuse 
the conclusive association of 
observed disturbances with dark-matter subhalo interactions.
Our own work is also motivated by these current observations which contain
tantalizing suggestions of non-uniformity in some cold stellar streams
\citep[e.g. the structure known as GD-1 and those from the globular cluster Pal 5, see][]{Odenkirchen2003a,Grillmair2006b,Grillmair2006c,Koposov2010a}, 
as well as the prospect of the density of
these streams being mapped more extensively (and accurately) in space
and velocity with observations in the near future. Such observations could
potentially distinguish between the effect of subhalo encounters on
different mass scales as well differentiate these signatures from
non-uniformities due to intrinsic stream dynamics.

We first review our understanding of the properties of dark-matter
structures and stellar stream evolution in smooth potentials in \S
\ref{back.sec}.  We use this understanding to make
order-of-magnitude estimates for the frequency and effect of encounters
of stellar streams with structures of different masses in \S
\ref{analytic.sec}.  We then go on to illustrate these expectations
with numerical experiments in \S \ref{numerical.sec} as well as
discuss the observational signatures of these encounters in 
\S \ref{discussion.sec}.
We summarize our conclusions in \S 6.

\section{Background and Methods}
\label{back.sec}

The aim of this paper is to characterize the effects that dark 
matter subhalos orbiting around the Milky Way can have on debris from 
satellite disruption. 
In order to isolate the influence of subhalos from other factors ,
neither the debris distribution nor the Milky Way is modelled self-consistently, and a simplified form for the mass distribution is assumed in both cases. 
In particular, in our numerical experiments, the debris is represented by 
3,000 test particles which orbit, along 
with a varying number of subhalos, in a smooth and 
spherical Milky Way halo. The orbits of the test particles and subhalos 
are integrated using the leap-frog method with a time-step of 0.5Myr.
The test particles respond to the gravity of both the Milky Way halo and 
subhalos, but the subhalos do not interact with each other. 

Note that our numerical approach misses several effects which could themselves 
contribute to non-uniform appearance of streams. These include multiple or continuous mass-loss, 
other asymmetries in the Galactic potential and self-gravity of the star streams. 
We will discuss each of these further in \S~\ref{confusion.sec}.

The dark-matter distributions (both parent and subhalos) 
are chosen to mimic the end point of the Via Lactea II \citep[VLII,][]{Diemand2008a}
N-body simulation of structure formation on Galactic scales (see \S \ref{dm.sec}),
while the distribution of tidal debris
(and test particle orbits in our numerical experiments) are chosen to match expectations 
from N-body simulations of satellite destruction (see \S \ref{streams.sec}). 

\subsection{Dark Matter Halo Properties}
\label{dm.sec}

The parent dark-matter distribution is a Milky-Way-like halo represented by a 
Navarro-Frenk-White (NFW) potential \citep{Navarro1996a}:
\begin{equation}
{\rho(r) \over \rho_{\rm crit}} = {\delta_{\rm c} \over (r/r_{\rm s})(1+r/r_{\rm s})^2}
\end{equation}
with parameters 
chosen to match the VLII N-body simulation 
\citep[$ M_{\rm MW}=1.77 \times 10^{12} {\rm M_{\odot}}, 
R_{\rm s}=24.6{\rm kpc}, R_{\rm vir}=389\rm kpc$,][]{Diemand2008a}.

\begin{figure}
\includegraphics[width=\columnwidth]{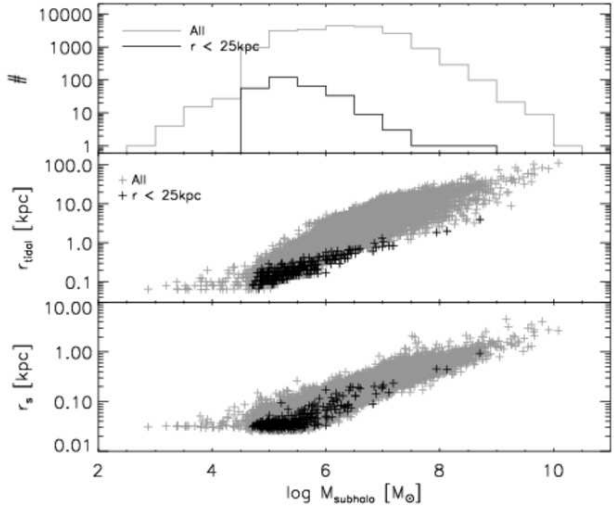}
\caption{Number, scale radii ($r_{\rm s}$) and tidal radii ($r_{\rm tidal}$) of subhalos in VLII as a function of subhalo mass ($M_{\rm sub}$). Gray points are for all subhalos and black are for those within 25 kpc at the present day.}
\label{vl.fig}
\end{figure}

We also assume the NFW form for the subhalos
and directly take all physical properties (i.e. the masses, tidal and scale radii, 
see Figure \ref{vl.fig}) and orbits of subhalos from the 
publicly available analysis of the VLII
simulation \citep{Diemand2008a}.
In practice, cosmological simulations like VLII have finite resolution,
so the mass function is not complete below 
$4 \times 10^6 {\rm M_{\odot}}$. 
We compensate for this resolution limit by duplicating
 subhalos within the mass range of 
$10^5 {\rm M_{\odot}}$ -- $10^7 {\rm M_{\odot}}$ to maintain the power-law 
mass spectrum with a power law index -1 set by larger masses (see upper panel of Figure \ref{vl.fig}). As a result, each mass decade has 10 times more subhalos than the mass decade above.
The orbits of the duplicated subhalos are set by
rotating the position and velocity vectors
of the original subhalos by random angles so as to preserve spherical symmetry 
and velocity isotropy of the subhalo population.
This method retains the potential and kinetic
energy of subhalos and their radial distribution.

\subsection{Tidal Stream Properties}
\label{streams.sec}

The general principle of debris evolution can be simply described:
stars being torn from a satellite initially share a single orbital
phase (or angle), but have a small range in orbital properties (or
actions); this range in orbital properties corresponds to a range in
orbital periods (or frequencies); and these period differences lead to
gradual spreading of the debris along the satellite's original orbit
\citep{Johnston1998a, Helmi1999a}.  Note that this description of evolution as simply 
phase-mixing can
break down for larger mass satellites (more than 0.01\% of the mass of
the parent) whose own gravity has been shown to influence the morphology of the
debris \citep{Choi2007a}. This effect is ignored in our work since we are
mostly interested in lower mass satellites and in differences in debris distribution
at smaller scales.

\subsubsection{Scales in Debris}

\citet{Johnston1998a} and \citet{Johnston2001a} found that the
distribution of debris particles observed in N-body
simulations of satellite disruption could be described
in terms of the dimensionless $tidal~scale$
\begin{equation}
  s \equiv \left({m \over M_{\rm p}}\right)^{1/3}
  \label{s.eqn}
\end{equation}
where $m$ is the mass of a satellite and $M_{\rm p}$ is the mass of 
a host halo enclosed by the pericenter, $R_{\rm p}$, of the satellite orbit.
The $tidal~radius$ where the internal and tidal forces are in equilibrium
and mass loss occurs scales as
\begin{equation}
  r_{\rm tide} \sim sR_{\rm p} 
           = \left({m \over M_{\rm p}}\right)^{1/3}R_{\rm p}.
  \label{eq2}
\end{equation}
The orbits of debris are offset in energy, $E$, and angular momentum $J$ from
the satellite's own orbital properties.
The tidal scales can be used to derive order-of-magnitude estimates for the characteristics sizes and ranges of these offsets, $\Delta E \sim \epsilon$ and $\Delta J \sim j$, where
\begin{equation}  
  \epsilon = r_{\rm tide} \left( {d\Phi \over dR}\right)_{R_{\rm p}} = s{GM_{\rm p} \over R_{\rm p}}
  \label{eps.eqn}
\end{equation}
and
\begin{equation}
  j = sJ.
  \label{j.eqn}
\end{equation}

Given these characteristic ranges, the angular length of streams of
debris as viewed from the Galactic center is expected to grow by of
order
\begin{equation}
\label{dpsi.eqn}
 	\Delta \Psi = \epsilon \left[ {2 \pi \over T_\Psi} {d T_\Psi \over d E} \right]_{J=J_{\rm circ}} ,
\end{equation} 
each orbit, where $J_{\rm circ}$ is the angular momentum of a circular
orbit of energy $E$ and azimuthal time-period $T_\Psi(E)$ at radius $R_{\rm circ}(E)$.
The secondary
dependance of orbital time periods on angular momentum has been
ignored.  The angular width is initially of order
\begin{equation}
\label{width.eqn}
	w=s.
\end{equation} 
and is expected to grow with time at a rate dependent on the parent
potential \citep{Helmi1999a}. For near-spherical potentials the rate is
sufficiently small that the approximation $w\sim s$ remains
useful for many orbits \citep{Johnston2001a}.

\subsection{The case of Palomar 5}

Tidal streams associated with the globular cluster Palomar 5 were
discovered using SDSS data by \citet{Odenkirchen2001a,Odenkirchen2003a} stretching several
degrees away either side of the center of the cluster, and have now
been mapped to a total extent of 22 degrees \citep{Grillmair2006b}.  They
were the first debris to be mapped from such a small
object and remain a primary example of a thin, cold stream.

\begin{table}
\begin{tabular}{|c|c|c|}
\hline
	Object			& Pal 5			& Sgr \\
\hline
\multicolumn{3 }{|c|}{Heliocentric view}	\\  
\hline
  $D_{\rm debris}$ (kpc)  & $\sim 23.5$ & 8-80 \\
  $w_{\rm obs}$ (degrees) & 0.5 & 10 \\ 
  $l_{\rm obs}$ (degrees) & 22 & $>$360 \\
\hline
\multicolumn{3 }{|c|}{Galactocentric view}	\\  
\hline
  $R_{\rm debris}$ (kpc)  & 18  & 8-80\\
   $l$ (degrees) & $>$ 29 & $>$ 360 \\
\hline
\hline
 \multicolumn{3}{|c|}{assumed properties} \\
\hline
 $R_{\rm p}$ (kpc) & 7.5 & 15.0 \\
 $R_{\rm a}$ (kpc) & 19.2 & 60.0 \\
 $T_R$ (Gyrs) & 0.33 & 0.89 \\
 $T_\Psi$ (Gyrs) & 0.55 & 1.35  \\
$R_{\rm circ}$ (kpc) &13.7 & 40.1 \\
$m_{\rm sat} (M_\odot)$ & $10^4$ & $5 \times 10^8$ \\
age (Gyrs) & 8.44  & 1.9 \\
 \hline
\multicolumn{3}{|c|}{derived scales} \\
\hline
$s$  & 0.007 &  0.17 \\
$\epsilon$ (km/s)$^2$ & 123 & 4572 \\
$\Delta \Psi$ (degrees) & 1 & 42 \\
$w$ (degrees) & 0.4 & 10 \\
\hline
\end{tabular}
\caption{Properties of the Pal 5 and Sgr streams.
\label{param.tab}
}
\end{table}
\vspace{5mm}

The first set of rows of Table \ref{param.tab} list the observed
characteristics of Pal 5's stream. Since Pal 5's orbit is not much
further from the Galactic center than the Sun, these are roughly
translated into a Galactocentric view in the second set of rows,
assuming the Sun is at 8 kpc from the Galactic center --- for example,
a lower limit on the Galactocentric angular length is taken to be
$l=(D_{\rm debris}/R_{\rm debris})l_{\rm obs}$ where $D_{\rm debris}$
and $R_{\rm debris}$ are the heliocentric and Galactocentric distances
respectively.  The observed properties can be broadly reproduced with
an N-body simulation of a hot, spherically symmetric system disrupting
along an orbit with characteristics listed in the third set 
of rows (``assumed properties'') of
Table \ref{param.tab}. The simulation adopted a 10,000-particle
Plummer model with mass $\rm 10^4 \msun$ and scale length 7.5 pc for Pal
5, and calculated its self-gravity using \citet{Hernquist1992a}
``self-consistent-field'' code. The model was allowed to evolve in the
NFW parent potential described in Section \ref{dm.sec} for 8.44 Gyrs.
Very similar masses and orbits were found with the more extensive
modeling by \citet{Dehnen2004a}.

\begin{figure}
 \begin{center}
  \vspace{5pt}
   \includegraphics[angle=90,width=\columnwidth]{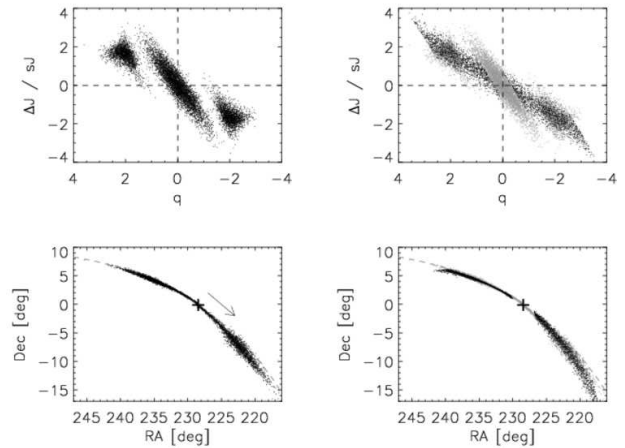}
   \caption{Different representations of the globular cluster, Pal
     5. N-body (left hand panels) and test particle streams (right hand
     panels) shown in energy/angular-momentum space (top panels) and
     in the sky (bottom panels).  The dashed lines in the top panels
     show the mean orbital properties adopted for the test-particle
     streams. In the right-hand panels, the gray dots repeat
     the N-body results from the left and the black dots are the test particles.  The dashed lines in the bottom
     panels represent the orbital path and the arrow shows the
     direction of the moving stream. }
 \label{scales.fig}
 \end{center}
\end{figure}

The left-hand panels of Figure \ref{scales.fig} summarize the results
of our simulation . The top left-hand panel shows the energy-angular
momentum distribution of debris particle, with axes scaled by the
factors given by equations (\ref{eps.eqn}) and (\ref{j.eqn}), and
listed in the bottom set of rows in Table \ref{param.tab}.  Note that
debris particles are systematically offset from those still-bound to the satellite, and distributed in the range $\pm 3 \epsilon$ and $\pm 3 j$
around the satellite's own orbital energy/angular-momentum --- a
result that is found to be largely independent of satellite mass,
scale, profile and orbit \citep[e.g.][]{Johnston1998a}.  Note that the
range in scaled-angular-momenta does depend (mildly) on the eccentricity of the
orbit, with debris on more eccentric orbits exploring larger ranges.

The bottom left-hand panel of Figure~\ref{scales.fig} shows the final
position of the particles projected onto the plane of the sky, as
viewed from a Solar position 
in our model. From our assumed mass and
orbit, our analytic estimate (i.e. equation \ref{width.eqn}) suggests an angular width $w\sim 0.4$ 
degrees for the debris.  While this is of the same
order as both the observed and simulated width, this prediction is
relatively insensitive to the assumed mass for the satellite, so the
agreement should only be seen as a weak confirmation that the mass of
the satellite is the correct order of magnitude.  

The estimated angular length of a stream of age $t$ viewed from the Galactic
center is
\begin{equation}
\label{length.eqn}
l=  4    \; {t \over T_\Psi} \; \left({R_{\rm circ} \over R_{\rm debris}} \right)^2 \; \Delta \Psi,
\end{equation}
where $\Delta \Psi$ is the expected angular growth per orbit calculated from equation (\ref{dpsi.eqn}) for the given mass and orbit. 
The additional scaling, dependent on the radial phase at which the debris is observed  (i.e. distance of debris from the Galactic center $R_{\rm debris}$), is included to account for the effect of debris density increasing/decreasing as the angular speeds decrease/increase ($\propto \left(1/R_{\rm debris} \right)^2$ in a spherical potential from conservation of angular momentum.
The extra factor of 4 reflects the range
that characterizes the width of the energy distribution seen in the top left-hand panel of Figure \ref{scales.fig}  (i.e. $\pm
2\epsilon$).
The prediction for the angular length of the stream, as
observed from the Solar position, is given by

\begin{equation}
\label{lobs.eqn}
  l_{\rm obs} =\left({R_{\rm debris} \over D_{\rm debris}}\right)  l.
\end{equation}

For the parameters adopted in our simulation of
Pal 5 we find $l_{\rm obs} \sim 30$ degrees, which agrees with the
length seen in the lower-left panel of Figure \ref{scales.fig}.

\subsubsection{Initial Conditions for Numerical Experiments}
\label{initialsetup.sec}

This paper employs a simplified representation of N-body results that
characterizes the evolution of tidal debris described above using test
particles integrated in the combined potential of the parent and
orbiting lumps, but ignoring the self-gravity of the satellite.  The
particles are given initial positions and velocities slightly offset
from the satellite's own, that are chosen to reproduce the scaled
energy and angular momentum distributions seen in the top left-hand
panel of Figure \ref{scales.fig}. Specifically, the particles are
initially positioned uniformly distributed along two lines pointing in
the direction of the Galactic center and centered on the satellite at
the apocenter of its orbit at Galactocentric radius $R=R_{\rm a}$ and
tangential velocity $v_{\rm tan}$.  Particles at position $\Delta R$
along the bars relative to the satellite are assigned a tangential
velocity $v_{\rm tan}+\Delta v$, where these offsets in position and
velocity satisfy the equations:
\begin{eqnarray}
	\Delta E &=& \alpha \epsilon = v_{\rm tan} \Delta v + \Delta R \Big( {d \Phi \over d R} \Big)_{R=R_{\rm a}}\cr
	\Delta J &=& \beta j = v_{\rm tan} \Delta R + R \Delta v.
\end{eqnarray}
Here $\alpha$ is restricted to the range $\pm 3$ and $\beta=2.5 \alpha/3$ in
order to mimic range in the orbital properties seen in the simulations
In addition, the particles are assigned
small random velocities in the $x$, $y$ and $z$ directions around the
mean $v_{\rm tan}+\Delta v$ chosen from a gaussian with width $\sigma=
0.6 s v_{\rm tan,apo}$ to reproduce the spread in angular momentum at
a given energy.

The right-hand panels of Figure \ref{scales.fig} illustrate the
success of this method by comparing energy/angular-momentum
distributions (top panels) and final positions (bottom panels) for our
N-body (gray particles) and test particle (black points) model of
debris evolution for Pal 5 (assumed properties in third set of rows of
Table \ref{param.tab}).

\section{Result I: Analytic Estimates}
\label{analytic.sec}

\begin{figure}
 \begin{center}
  \vspace{5pt}
   \includegraphics[angle=0,width=\columnwidth]{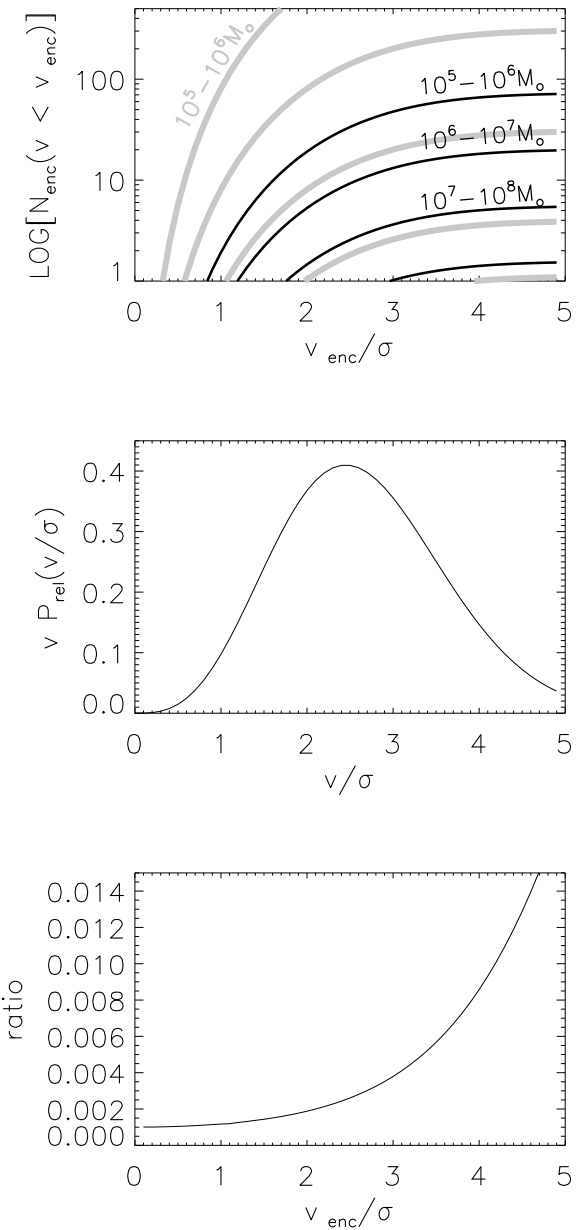}  
   \caption{Encounter frequencies. The upper panel plots the estimated number of direct encounters for Pal 5 (solid lines) and Sgr (gray lines) debris during the debris lifetime and for subhalos of different masses. Each line presents the case for different subhalo mass decades as noted on the lines. The middle panel shows the distribution of encounter velocities. The bottom panel shows an estimate for the fraction of encounters of fixed energy that are due to slow encounters with small masses (see text).}
 \label{freq.fig}
 \end{center}
\end{figure}

In this section we use simple analytic arguments to derive
order-of-magnitude estimates for the frequency of encounters of streams
with varying mass dark-matter substructures (\S \ref{freq.sec}) and
the characteristic scales of these subhalos' influence on the debris
(\S \ref{enc.sec}).  We first illustrate the implications of these
estimates for the debris from Pal 5 and then go on to discuss
the application to other streams (\S \ref{other.sec}).

\subsection{Frequency of Encounters}
\label{freq.sec}

Suppose a (perfectly tubular!) stellar stream of angular length $l$ and
angular width $w$ is orbiting at an average distance $R_{\rm
  circ}$ from the Galactic center through a sea of dark-matter
subhalos of size $r_{\rm tidal}$ and number density $\nl$,
 The rate of encounters that directly impact the stream with   
relative speeds in the range $(v, v+dv)$ can be estimated as:
\begin{eqnarray}
\label{rate.eqn}
	{d \nenc \over dt}
&=& {\rm volume \; encounter \; rate} \times {\rm number \; density} \times P_{\rm rel} (v) dv	\cr
&=&({\rm stream \; surface \; area})\times v \times \nl \times P_{\rm rel} (v)dv\cr
&=& (R_{\rm circ}\; l \times 2 \pi  b_{\rm max} )\times \nl \times  v\;  P_{\rm rel} (v) dv.
\end{eqnarray}
where $P_{\rm rel}(v)$ is the probability distribution of relative speeds.
The maximum impact parameter for direct encounters is taken from the condition that the stream and subhalo overlap physically:

\begin{equation} 
b_{\rm max}={\rm MAX}[wR_{\rm circ}/2,r_{\rm tidal}].
\end{equation}
Assuming that the local
velocity distribution of the subhalos is an isotropic Maxwellian with dispersion
$\sigma$, we can 
approximately represent the relative distribution of encounter speeds along a typical orbit by
the relative speed distribution between the subhalos

\begin{equation}
 P_{\rm rel} (v) dv= {1 \over (\sqrt{2\pi} \sigma)^3} v^2 \exp\left(-{v^2 \over 4 \sigma^2}\right)dv
\end{equation}
\citep[see][equation (8.46)]{Binney2008b}.
This implies a typical encounter speed of
\begin{equation}
v_{\rm enc}^{\rm typ}={\int_0^\infty v^2 P_{\rm rel}(v) dv \over \int_0^\infty v P_{\rm rel} (v) dv }= 3\sqrt{\pi}\sigma/2.
\end{equation}
Replacing $l$ in equation (\ref{rate.eqn}) with equation (\ref{length.eqn}) (evaluated for $R_{\rm debris}=R_{\rm circ}$)
and integrating  
over speeds and time, we find
the number of encounters with encounter speeds less than $v_{\rm enc}$
that this stream experiences over its lifetime $t$:
\begin{eqnarray}
\label{nenc.eqn}
	\nenc(v< v_{\rm enc})
&=& 8 \sqrt{\pi} R_{\rm circ}\; b_{\rm max} \sigma t \; \nl \times  \Delta \Psi   \left({t \over T_\Psi}\right) \cr
&&\times  \left[ 1 - \left( 1 +{v_{\rm enc}^2 \over 4 \sigma^2}\right) \exp\left(-{v^2_{\rm enc} \over 4 \sigma^2}\right)\right] .
\end{eqnarray}

The solid lines in the top panel of Figure \ref{freq.fig} show 
$\nenc(v< v_{\rm enc})$ for our Pal-5-like stream (i.e. evaluating
equation \ref{nenc.eqn} with parameters adopted from Table
\ref{param.tab}) as a function $v_{\rm enc}$
for different decades in subhalo masses and assuming 
a characteristic $\sigma$ of 120 km/s. 
(Note that the circular velocity at radius 25 kpc in a spherical NFW halo with parameters chosen to match the parent halo in VLII is 178 km/s, which suggests a dispersion of 178/$\sqrt{2}$=126 km/s.)
The subhalos in each group
are assumed to have tidal radii similar to those seen in VL II for subhalos within 25kpc
(i.e. $r_{\rm tidal} = 0.21, 0.58, 1.6, 4.5, 12.7$kpc for masses in
ranges $10^5-10^6/10^6-10^7/10^7-10^8/10^8-10^9/10^9-10^{10} \rm M_\odot$
respectively --- see Figure \ref{vl.fig}).  
For $\nl$ we adopt the mean number density within 25 kpc 
given by assuming there to be 42  subhalos in the range
$10^6-10^7 M_\odot$ within this volume (as seen in VLII, see Figure \ref{vl.fig}), and a factor
of 10 more/less in each subhalo mass decade below/above (to mimic the mass spectrum seen in VLII).  
(This is expected to be an upper limit on the number of encounters since the adopted age produces a stream slightly longer than the observed stream, and the mass adopted for our Pal 5 model was fairly low.)
Inserting these numbers for the $10^6-10^7 M_\odot$  mass range explicitly into equation \ref{nenc.eqn} and letting $v_{\rm enc} \rightarrow \infty$ gives the total number of direct encounters:

\begin{eqnarray}
\label{numbers.eqn}
\lefteqn { N_{\rm enc, total} (\ml=10^6-10^7 M_\odot) = 
20 \;  \left(R_{\rm circ}\over 13.7 {\rm kpc}\right) {}} \nonumber\\
& & \times \; \left(b_{\rm max}\over 0.58 {\rm kpc}\right)
\; \left( \sigma \over 120 {\rm km/s}\right) 
\; \left(t \over 8.44 {\rm Gyrs}\right)^2 \cr
& & \times  \; \left(\nl \over 0.0006 {\rm kpc}^{-3}\right)
 \left(\Delta \Psi \over 1^\circ\right)  
 \left({0.55 {\rm Gyrs}\over T_\Psi}\right) \cr
\end{eqnarray}

Overall, our calculations indicates that Pal 5's stream will have suffered hundreds of direct encounters with subhalos less massive than $10^5 \rm M_\odot$, $\sim 70$ with subhalos masses in the range $10^5-10^6 \rm M_\odot$, $\sim 20$ with subhalos of $10^6-10^7 \rm M._\odot$, $\sim 5$ with subhalos of $10^7-10^8 \rm M_\odot$, and a few with subhalos of $10^8-10^9 \rm M_\odot$. 
Subhalos in the higher mass bins are very unlikely to
directly hit the stream, although they will influence it through more
distance encounters.  In addition, note that while there are
sufficient numbers of smaller subhalos to fully explore the relative
velocity distribution, the few encounters with large subhalos are
likely to take place close to the typical encounter speed around the
peak of the velocity distribution shown in the second panel.

\subsection{Effect of Encounters}
\label{enc.sec}

In order to develop some understanding of the effect of subhalos on streams, consider the idealized case of a subhalo encountering a perfectly straight stellar stream
aligned with the $x$-axis, with impact parameter $b$ along the
$y$-axis and relative velocity ${\bf v_{\rm enc}}=(v_x,0,v_z)$.  
Overall, the
change in speed of the stars perpendicular to the stream is zero by
symmetry. Assuming the change in the relative speed for stars in the
stream is negligible during the encounter ( $\Delta v_x << v_x$) and
simply integrating the equations of motion we find
\begin{eqnarray}
\label{dv.eqn}
\Delta v_x &=& \int_{-\infty}^{\infty} a_x dt \nonumber \\
 &=& \int_{-\infty}^{\infty} {G M_{\rm sub} (v_x t+ x) \over [ (v_x t+x)^2 +b^2 +v_z^2 t^2]^{3/2}} dt \nonumber \\
 &=& -{2 G M_{\rm sub} x \over  v_{\rm enc} b^2} {v_z^2 \over v_z^2 (x/b)^2 +v_{\rm enc}^2}
\end{eqnarray}
where $x$ is the position of the star in the stream at the moment of
closest approach of the subhalo, relative to the impact point.  If the
space motion of the stars in the streams in the Galactic rest-frame is
$v_{\rm stream}$, then the energy change is
\begin{equation}
    \Delta E  = {\bf v}_{\rm stream} \cdot \Delta {\bf v} + {1\over2} \Delta {v}^2 = v_{\rm stream} \Delta v_x +O(\Delta v_x^2)
  \label{de.eqn}
\end{equation}
For direct  encounters with subhalos of mass $M_{\rm sub}$ :
\begin{equation}
\label{de0.eqn}
\Delta E (b=0) = 2 {G M_{\rm sub}\over x} {v_{\rm stream} \over v_{\rm enc}}.
\end{equation}

\subsubsection{Lowest Mass Subhalos}
\label{lowmass.sec}

Figure \ref{freq.fig} suggests Pal 5 experiences thousands of encounters with subhalos less massive
than $10^5 \rm M_\odot$  during its lifetime.
In this many, weak encounter regime the first term in equation
(\ref{de.eqn}) should average out to zero and leave the second term to
heat the stellar stream.  Despite their large number, since the
heating term is second order, these encounters would have negligible
effect compared to the few encounters with larger subhalos. For
example, while there would be a factor of ten more encounters with
$M_{\rm sub} \sim 10^4-10^5 \rm M_\odot$ than in the mass decade above, the net
energy change due to each of those encounters would be a factor of 100 smaller, so the individual encounters with larger lumps would be generally be expected to have a more significant effect.

Nevertheless, equation (\ref{de0.eqn}) indicates that the effect of a lower
mass subhalo can be comparable to that of a subhalo in the mass decade above if the encounter with the lower mass is 
$\sim$ 10 times slower. Indeed, the large number of encounters with low mass halos ensures that a small number will come from the low-end stream of the relative velocity distribution (see middle panel of Figure \ref{freq.fig}) and the effect of these few, slow encounters would not be expected to average out.
The lower panel of Figure \ref{freq.fig}
compares the frequency of these {\it slow} encounters with low mass
subhalos relative to ones of comparable influence in the mass decade
above by
plotting the ratio of the number of encounters of low mass subhalos to higher mass subhalos of comparable influence: $[10 \nenc(v< v_{\rm enc}/10)]/\nenc(v<v_{\rm enc})$. 
Since this ratio is never much greater than 1\% we conclude that
while rare, influential low-speed encounters with low mass subhalos
can occur, they will be much less frequent than encounters of
comparable influence with higher-mass
subhalos.

\subsubsection{Intermediate Mass Subhalos}
\label{dir.enc.sec}

\begin{figure}
 \begin{center}
  \vspace{5pt}
   \includegraphics[angle=0,width=\columnwidth]{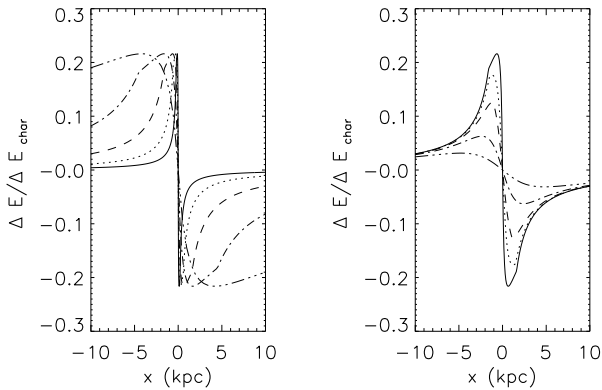} 
   \caption{Estimate for the scaled energy imparted by a direct  encounter at $x=0$ at different points along a stream. The left hand panel shows the influence of subhalos of mass $5\times 10^5/5\times 10^6/5\times 10^7/5 \times 10^8/5 \times 10^9 \rm \msun$. The right hand panel plots the result for a subhalo of mass $5 \times 10^7 \rm \msun$, traveling perpendicularly to the stream  and with impact parameters  $b=0/2/4/8/16 r_{\rm s}$.}
 \label{delE.fig}
 \end{center}
\end{figure}

Pal 5 should have experienced $\sim$20 direct encounters with
subhalos in the mass range $10^6-10^7 \rm \msun$, whose influence is
unlikely to average to zero. In order to gain some intuition
for their effect, the left hand panel of Figure \ref{delE.fig} plots
equation (\ref{de0.eqn}) for values for the stream and encounter speeds averaged over their respective velocity distributions
($v_{\rm stream}^{\rm typ}=2\sqrt{2/\pi} \sigma$ and $v_{\rm enc}^{\rm typ}=3\sqrt{\pi}\sigma/2.$) and $M_{\rm sub} = 5\times 10^5/5\times 10^6/5\times 10^7/5 \times 10^8/5 \times 10^9 \rm \msun$ (solid...dot-dot-dot-dash lines), as a
function of position $x$ along the stream.  The extended nature of
these subhalos was taken approximately into account by replacing $\ml$
in equation (\ref{dv.eqn}) with the mass enclosed within radius
$r=x$ for an NFW model with scale $\rs$ ($M_{\rm sub}(r)$).  (Note that the
integral was solved analytically and did not take this into account).
The subhalos had $\rs$ chosen to match those within 25kpc in VLII
($r_s=0.05,0.12,0.30,0.76,1.94$kpc respectively). In order to show the
behavior of $\Delta E$ for these different mass ranges in a single
plot, the energy change was normalized by the characteristic energy for a direct encounter, calculated by substituting $x=\rs$ in equation (\ref{de0.eqn}): 
\begin{equation}
\label{dechar.eqn}
	\Delta E_{\rm char}= 2 {G M_{\rm sub} \over \rs} {v_{\rm stream}^{\rm typ} \over v_{\rm enc}^{\rm typ}} = {8\sqrt{2} \over 3 \pi} {G M_{\rm sub} \over \rs}.
\end{equation}	 
Most noticeable in Figure
\ref{delE.fig} is the abrupt sign-change in $\Delta E$ at the impact
point --- stars upstream of the impact point are pushed to orbits of
higher energy and stars downstream of the impact point are pushed in
the opposite direction to orbits of lower energy. This effect
exaggerates the energy gradient already present along the stream (see
\S \ref{streams.sec}) and creates a gap in energy.  The fraction of the
stream affected by the encounter depends on the size of the perturber
--- with the smallest subhalos influencing only a subset of stream
particles directly around the impact point (solid lines) and the
largest subhalos influencing the whole stream (triple-dot-dashed
line).

\begin{figure}
 \begin{center}
  \vspace{5pt}
   \includegraphics[angle=0,width=\columnwidth]{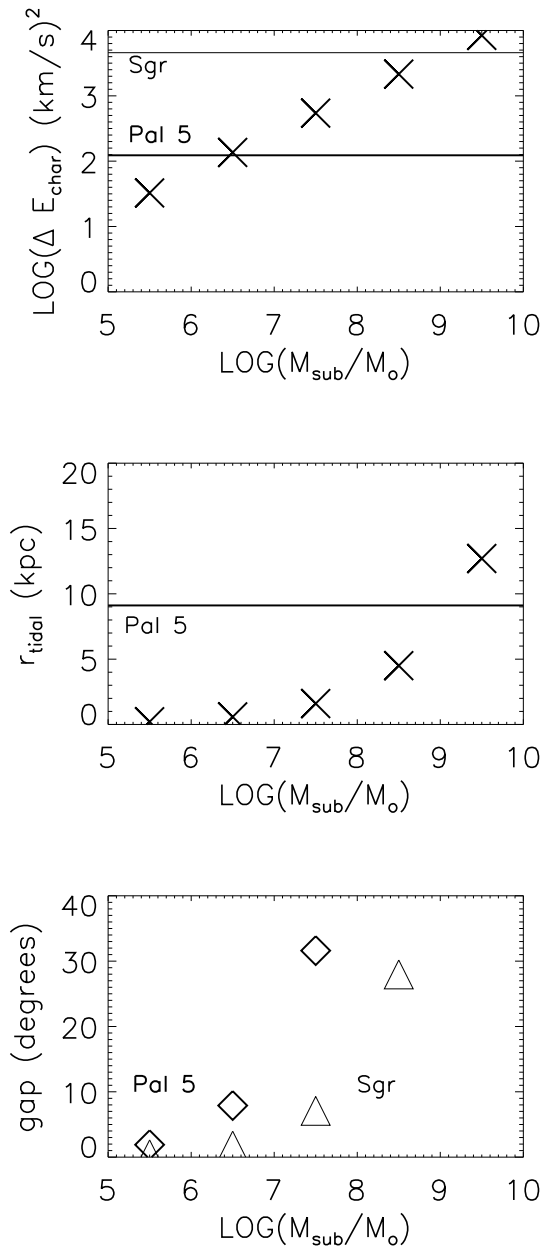} 
   \caption{Scales in an encounter relative to stream properties for subhalos of different masses. The top panel compares $\Delta E_{\rm char}$ (symbols) with the intrinsic energy spread in Pal 5's (bold line in the middle) and Sgr's (faint line on the top) debris. The middle panel compares the length of Pal 5's stream with the tidal radius of the subhalos (Sgr's stream is far longer than the plot). The lower panel gives an upper limit on the angular scales of gaps due to encounters with Pal 5 (bold diamond symbols) and Sgr (faint triangle symbols) debris. }
 \label{encscales.fig}
 \end{center}
\end{figure}

How influential these encounters are on debris evolution also depends on stream characteristics. The top and middle panels of Figure \ref{encscales.fig} assess this by plotting $\Delta E_{\rm char}$ for subhalos of different masses and the spatial scale over which this energy change is important (taken to be the tidal radius $r_{\rm tidal}$). For comparison, the energy scale over which debris orbits are spread for Pal 5 (i.e. $\epsilon$, calculated from equation \ref{eps.eqn}) and the physical length of the stream itself are shown as bold lines.
These plots suggest that each direct encounter of Pal 5's debris with subhalo masses in the range $10^6-10^7 \rm \msun$ should create gaps in energy that are a significant fraction of the spread of energy already present in the debris (i.e. $\Delta E_{\rm char} \sim \epsilon$) for of order 10\% of stars in the stream (i.e. $r_{\rm tidal} \sim l/10$).

Over time these gaps in energy will cause the stream stars around the impact point to spread apart from each other faster than stars in other regions of the stream, leading to physical gaps in the stream density distribution. We can estimate an upper limit on the angular scale of these physical gaps by replacing the stream energy range $4\epsilon$ in equation (\ref{length.eqn}) with $\Delta E_{\rm char}$ and calculating $l_{\rm obs}$ for Pal 5's lifetime and orbit (assumed parameters in Table 1). The bold diamonds in the lower panel of Figure \ref{encscales.fig} show this estimate. 
 
Overall, we conclude that direct encounters of Pal 5's debris with subhalo masses in the range $10^6-10^7 \rm \msun$ during its lifetime could lead to about 20 gaps in density of order several degrees across, while masses in the range $10^5-10^6 \rm \msun$ could leave many more on sub-degree scale.
Few, if any,  direct encounters are likely to occur for the Pal 5 stellar stream for masses greater than $\sim 10^7 \rm \msun$ (as shown in Figure \ref{freq.fig}), but could potentially lead to larger gaps.

\subsubsection{Indirect Encounters with Largest Mass Subhalos}
\label{indir.enc.sec}

The right hand panel of Figure \ref{delE.fig} repeats the same plot 
as the left hand panel, but this time for a $10^7 \rm \msun$ subhalo traveling perpendicularly to the stream (i.e. $v_{\rm enc}=v_{\rm enc}^{\rm typ}=v_z$) and
passing it at impact parameters $b$ of 0,2,4,8,16 $\rs$. 
Note that for these more distant encounters the impulsive
approximation under which equation (\ref{dv.eqn}) was derived breaks down
and the estimate is likely to overestimate the size of a subhalos' influence.
The results are nevertheless included here as a broad guide to general behavior,
but should be treated with some skepticism.
The effect of more distant encounters can be assessed more realistically using numerical 
integration (see \S \ref{numerical.sec}).

As the impact parameter increases, both the size of the energy change 
decreases and the scale over which it is felt grows. The energy imparted to the stream changes more smoothly as a function of position along the stream than in the direct encounter case, with the most extreme changes apparent at the edges. This is in part because the impact parameter is now comparable to the length of the stream itself and it is only at the edges that the velocity change is aligned with the stream velocity to produce maximum energy change. The net result is that the stream is smoothly stretched in energy, rather than forming an abrupt gap.
It is this type of interaction that dominates for the larger masses: Figure \ref{freq.fig} tells us we should expect few such encounters over Pal 5's lifetime.

Note that the same behavior can be seen for individual encounters for smaller masses, but in those cases there are sufficient numbers of weak encounters for the first-order terms in energy change to cancel out. So we conclude that indirect encounters of smaller masses are not important for leaving observable signatures.

\subsection{Summary and Application to Other Streams}
\label{other.sec}

In summary, our analytic estimates indicate that Pal 5's stream should
be sensitive to dark subhalos, with unique signatures from different
subhalo masses.  The dominant influence on streams like Pal 5's  will be from
scattering of stars by encounters with subhalos in the mass decades
$10^5-10^7 \rm\msun$ where many tens of encounters occur
during the stream's lifetime.  These events will lead to gaps in the
stream with characteristics scales reflecting the mass of the
perturbers.  Low-speed encounters from smaller subhalos can create
comparable gaps, but these events occur much less frequently and so
should not confuse the signal.  Heating by perturbers in the lower
mass decades should also be negligible.  Direct encounters with larger
masses are unlikely (though catastrophic if they occur), and indirect
encounters will tend to stretch rather than chop up a stream.

Our estimates also suggest that streams with Pal 5's scales are the
most interesting.  The thin lines and symbols in Figure \ref{freq.fig}
illustrate the results of repeating the calculations made above for
Pal 5's stream for Sgr's debris.  In comparison to Pal 5, these
streams have a far larger cross section for encountering dark-matter
subhalos, but also have a far larger spread in orbital properties in
which to hide the signatures of encounters.  This means that the
influence of encounters with subhalos less massive than $10^8 \rm\msun$
are likely to be negligible as: (i) both the spatial and energy scales
of the change are much smaller than the inherent stream scales (see
first and second panels of figure \ref{encscales.fig}), and (ii) the
encounters become frequent enough that first order changes may be
expected to cancel out.  Subhalos more massive than $10^8 \rm\msun$
could leave their imprint on Sgr debris.  However, these dark-matter
structures are generally thought to contain stars and hence should be
detectable by other means.

We conclude that for the purposes of finding {\it missing} satellites,
cold streams from low mass objects should be the most sensitive
probes.

\section{Result II: Numerical Illustrations}
\label{numerical.sec}

This section presents the results of scattering test particle streams by subhalos
with numerical integration.
We present our results in orbital (i.e. energy/angular-momentum) space, 
coordinate space, and observable space. The Sun is assumed to be at 
(X,Y,Z) = (-8kpc, 0, 0).

In the following subsections, we look first at individual close encounters
(\S \ref{individual.sec}), then multiple encounters for subhalos in each
mass decade (\S \ref{multi.sec}), and
finally examine the combined effect of all subhalos in the full  $\rm \Lambda CDM$ 
mass spectrum(\S~\ref{lcdm.sec}).

\subsection{Individual Encounters}
\label{individual.sec}

\begin{figure*}
 \begin{center}
  \vspace{5pt}
   \includegraphics[angle=90,width=2\columnwidth]{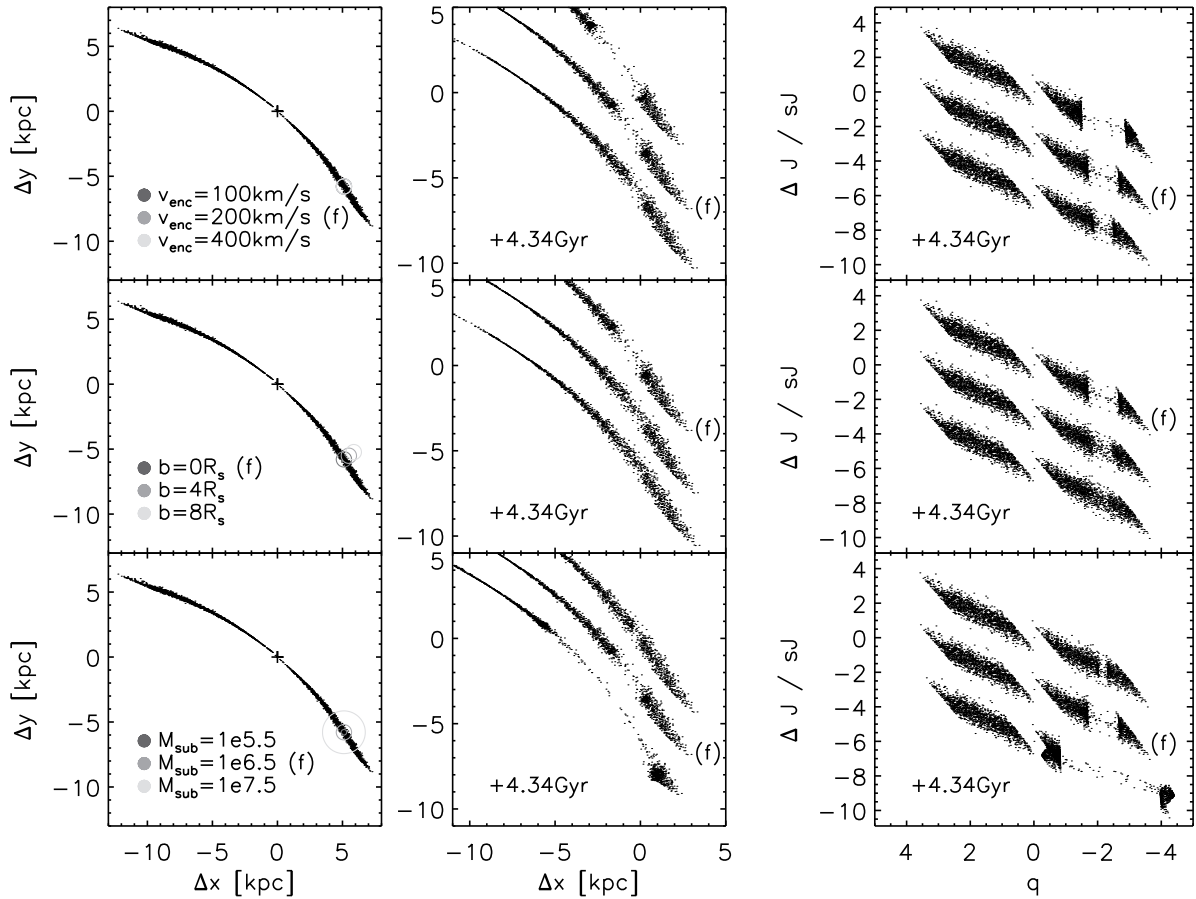} 
   \caption{The various effects of encounters on a Pal 5-like stream. 
   Our fiducial encounter (particles labelled {\it (f)} in each panel)
   has $M_{\rm sub} = 10^{6.5} \rm \msun$, $v_{\rm enc}=200 \rm km/s$, $b=0$.
   The top/middle/bottom rows show the responses when the velocity/impact/mass are
   allowed to vary around this. 
    The ``$+$'' in each panel represents the Pal 5 center, and the circles show the size ($r_{\rm tidal}$) of the encounters and where they pass through. Note that three encounters are all overlapped in the top-left panel.
   The streams in the first, second, and last columns are
   centered at the Pal 5 center, impact point, and mean energy/angular-momentum respectively. Streams other than the fiducial model are shifted by -3 and -6 along y-axis from their original positions in the second and last columns.
   }
 \label{ind.fig}
 \end{center}
\end{figure*}

We first follow a single artificial encounter with $M_{\rm sub}=10^{6.5} {\rm M_{\odot}}$,
$r_{\rm s}=0.118{\rm kpc}$ and $r_{\rm tidal}=0.590{\rm kpc}$ passing directly
through a Pal 5-like stream perpendicular to the orbital plane with a relative 
velocity $(v_{\rm x}, v_{\rm y}, v_{\rm z}) = (0, 0, -200{\rm km/s})$ to the stream.
Figure~\ref{ind.fig} illustrate the effect of this
direct encounter on the stream (fiducial results in each panel labelled with {\it (f)}). 
The orbital plane of the stream is in
the X-Y plane. The left column shows
the moment of closest approach of the subhalo encounter moving along the
z-direction and passing perpendicularly through the stream. 
The middle and right columns show the particle positions in the X-Y  and 
energy/angular-momentum planes
4.34 Gyr after the encounter.
There is a clear energy gap around the impact point
in the stream that forms immediately after closest approach:
 the particles behind the impact points (with
higher energy) are accelerated (gain energy) and the ones ahead the
impact points (with lower energy) are decelerated (lose energy) as
predicted in \S \ref{dir.enc.sec}. 
This energy gap results in a spatial gap along the
stream which grows with time due to phase mixing 
as illustrated in the middle column. 
Hence, a stream influenced by many encounters
should end up having a clumpy structure  
produced by numerous energy and spatial gaps.

Having illustrated the nature of the gaps in energy, we now investigate how these
response of the stream to varied encountering conditions.
The first row of Figure~\ref{ind.fig} illustrates the effect of
encounters with different velocities relative to the stream 
($v_{\rm enc}=100, 200, 400 {\rm km/s}$), the second row shows the gaps
resulting from encounters passing with different impact parameters
($b_{\rm impact}=0\times, 4\times, 8\times r_{\rm s}$ which correspond
to 0, 0.470, 0.940kpc), and the last row shows various mass encounters
($M_{\rm sub} = 10^{5.5}, 10^{6.5}, 10^{7.5} {\rm \msun}$). Overall, when the
velocity of the encounter doubles the gap becomes half, while smaller
impact parameters and larger masses result in larger energy
gaps. These scalings agree with the trends in the 
equations (\ref{dv.eqn}), (\ref{de.eqn}),
and (\ref{de0.eqn}).  For a stellar stream orbiting with many dark
matter subhalos, irregular clumps are expected in the stream that
correspond to various energy gaps created by different encounters at different times.

Lastly, the bottom panels suggest that while a small subhalo has an effect
on the local region of a stream generating a small energy gap, a close
encounter with a subhalo larger than $\rm 10^{7.5} {\rm \msun}$
can globally distort the entire energy-angular
momentum scales, since the size and influence of the subhalos are larger than the
entire stream. This will be discussed in more detail in \S~\ref{multi.sec}.

\subsection{Multiple Encounters}
\label{multi.sec}

\begin{figure}
 \begin{center}
  \vspace{5pt}
   \includegraphics[width=\columnwidth]{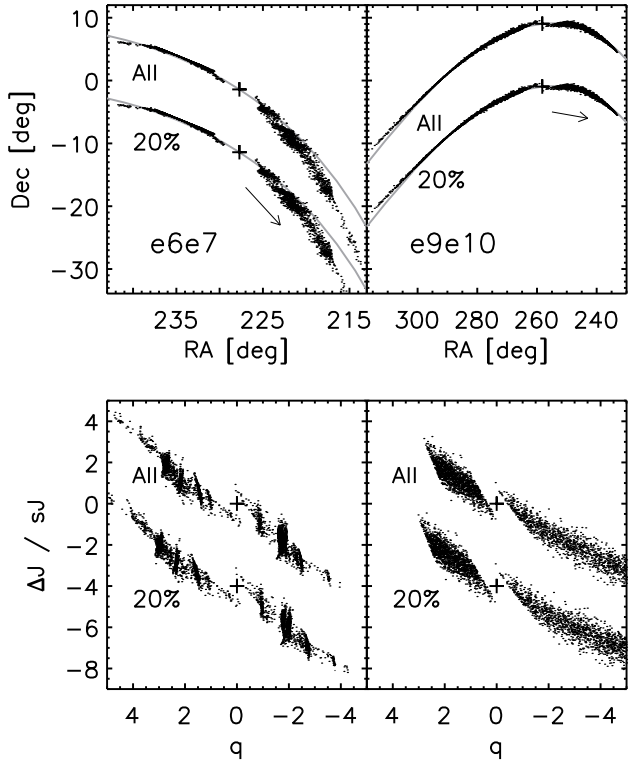} 
   \caption{The final distribution of stream particles from simulations of encounters with 
   all the subhalos (upper particle distributions) and the 20\% closest encounters (lower particle distributions) in the mass range 
   $10^6-10^7{\rm \msun}$ (left panels) and $10^9-10^{10}{\rm \msun}$ (right panels).
   The top row shows the stream in the sky and the bottom row 
   shows the energy/angular-momentum distribution. The arrows illustrate the direction of 
   the stream movement and the cross represents the center of host satellite Pal 5. 
   The grey lines illustrate the orbits of satellites.
   The lower particles have been shifted for clarity
   by -$10^{\circ}$ and by -4 along the y-axis in the top and bottom panels respectively.}
 \label{close.fig}
 \end{center}
\end{figure}

To examine the integrated influences of many subhalos, we perform separate
simulations with subhalos drawn from each decade of the $\rm \Lambda CDM$ mass
spectrum.
We first illustrate our results by contrasting simulations with the $\sim$30,000
intermediate mass subhalos in the mass range $10^6 - 10^7\rm \msun$ 
and $\sim$30 large mass subhalos of $10^9 - 10^{10}\rm \msun$.
Figure~\ref{close.fig} shows the final spatial and 
energy/angular-momentum distribution of these cases. 
First consider the overall morphologies in Figure~\ref{close.fig}.
Recall that the calculation in \S\ref{freq.sec} suggests there should be $\sim 10$ direct
encounters with subhalos in the lower mass range, which is in rough agreement with the visual impression of the 
energy/angular-momentum distribution in the lower left-hand panel.
In contrast, the stream with the large subhalos (the right column of Figure~\ref{close.fig})
does not have any gaps. Instead, the orbital phase of the stream with 
subhalos in the mass range
$10^9 - 10^{10}\rm \msun$ is shifted and the entire energy scale is
changed as suggested in \S~\ref{individual.sec}.
All panels in Figure~\ref{close.fig} are plotted relative to a central particle (marked with +). The scaled energy $q$ and angular momentum
($\Delta J / sJ$) values of the central particles changed by (-0.74, -0.53)
and (9.40, 4.00) for the streams with $10^6 - 10^7{\rm \msun}$ and 
$10^9 - 10^{10}\rm \msun$ subhalos respectively.  
This phase and energy shift will not be observable
since there is no way to probe the original phase and energy.

In order to disentangle the effect of direct (close) encounters from indirect (distant) ones,
we select the 20\% closest encounters based on the minimum distance between the subhalos 
and the stream particles  that occurs throughout the simulations. We 
then re-run the same simulation with only these subhalos present.
The results, presented in the lower particle distributions in each panel of Figure~\ref{close.fig}, 
are almost identical to the ones with all subhalos in each mass range, which suggest that 
heating by the more distant encounters  is 
negligible.

\begin{figure}
 \begin{center}
  \vspace{5pt}
   \includegraphics[width=\columnwidth]{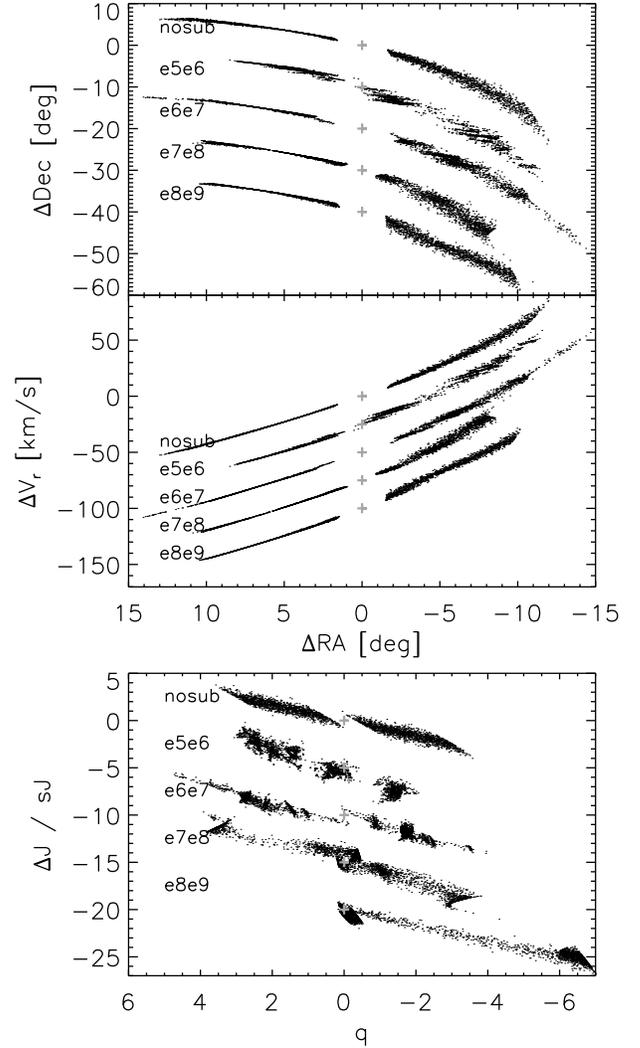} 
   \caption{Final distribution of stream particles from simulations with subhalo 
   encounters of varying mass ranges.
   The streams with no subhalos, subhalos in the mass range $10^5-10^6{\rm \msun}$,
   $10^6-10^7{\rm \msun}$, $10^7-10^8{\rm \msun}$, and $10^8-10^{9}{\rm \msun}$
   are shown in the sky (top), radial velocity (middle), and energy/angular-momentum space (bottom). 
   The stream with no subhalos is at the original position and the others are 
   shifted down by 10 and 5 for the top and bottom panel respectively. The grey crosses
   represent the center of the host satellite.}
 \label{varyingmass.fig}
 \end{center}
\end{figure}

Figure~\ref{varyingmass.fig} repeats the plots in Figure~\ref{close.fig} for 
simulations with subhalos in all mass decades. While the stream without any subhalos 
shows a smooth morphology in the sky and 
energy/angular-momentum space, the streams with subhalos in the mass range
$10^5-10^6 {\rm \msun}$, $10^6-10^7 {\rm \msun}$, and $10^7-10^8 {\rm \msun}$
show eye-catching signatures such as spatial, velocity, and energy gaps. As expected,
the smaller subhalos cause many more smaller scale clumps
in the stream than the large ones. In contrast, the stream with
$10^8-10^9 {\rm \msun}$ subhalos does not have clear signatures.
These subhalos are too large to leave small scale variations but too small to shift 
the entire stream as in Figure~\ref{close.fig}. Rather, these subhalos distort 
the energy-angular momentum distribution and slightly change the overall shape of the stream.

The gaps in Figure~\ref{varyingmass.fig} due to $10^5-10^6 {\rm \msun}$
subhalos are more commonly smaller than those due to the $10^6-10^7 {\rm \msun}$ objects.
This indicates that each mass decade might be separately detected.
While small encounters with subhalos will generally result in smaller energy 
gaps which correspond to smaller physical gap than encounters with large subhalos, note that
in the middle of the leading part of the stream with $10^5-10^6 {\rm \msun}$ subhalos, 
there is a large gap (where $\Delta RA \sim -5^\circ$) which is as big as the ones 
in the stream with $10^6-10^7 {\rm \msun}$ subhalos. On further investigation,
we found this large gap can be attributed to a very slow encounter 
(which caused a large gap in energy) that occurred relatively early in the simulations
(which allowed the gap to grow). However, we expect this to be a rare occurrence
as described in \S \ref{freq.sec}.
To confirm this expectation, we ran four additional simulations with the same 
conditions as the original $10^5-10^6 {\rm \msun}$ run but different initial 
starting points. In none of these cases did such large gaps appear. 

We can also extrapolate from
these numerical results to consider the effect of the smallest {``missing satellites''} with 
$\ml < 10^5 \rm\msun$. The scales of the gaps due to individual encounters with these subhalos will be even smaller than those due the mass decade above (like the ones apparent in Figure~\ref{varyingmass.fig}) and would thus require a very high resolution map to detect.
Making such a map is observationally challenging since Pal 5's stream contains only a similar number of stars to the globular cluster itself (i.e. of order 1,000 stars), and at much lower
surface density compared to the background field population.
Moreover, as argued in \S~\ref{lowmass.sec} and also apparent in 
Figure \ref{varyingmass.fig},
the averaged heating of many such encounters is negligible compared to responses to higher mass subhalos.
We conclude that it is unlikely that the smaller subhalos would leave an observable signature in Pal 5's stellar streams.

\subsection{Stellar Streams in $\Lambda$CDM}
\label{lcdm.sec}

To assess the integrated effect of subhalo encounters, we would ideally
run simulations of streams evolving in the presence of dark-matter
subhalos from the full $\Lambda$CDM spectrum.  However, as outlined in
\S \ref{multi.sec}, close encounters play a decisive role in shaping
streams.  Hence, we limit our computational expense by selecting only
the 97,090 subhalos which enter within a Galactocentric radius of
25kpc at least once during the 8.44Gyr fiducial simulation time from our full
realization of over 300,000 subhalos in the mass range $10^5 - 10^{10}\rm
\msun$.

\begin{figure*}
 \begin{center}
  \vspace{5pt}
   \includegraphics[width=2\columnwidth]{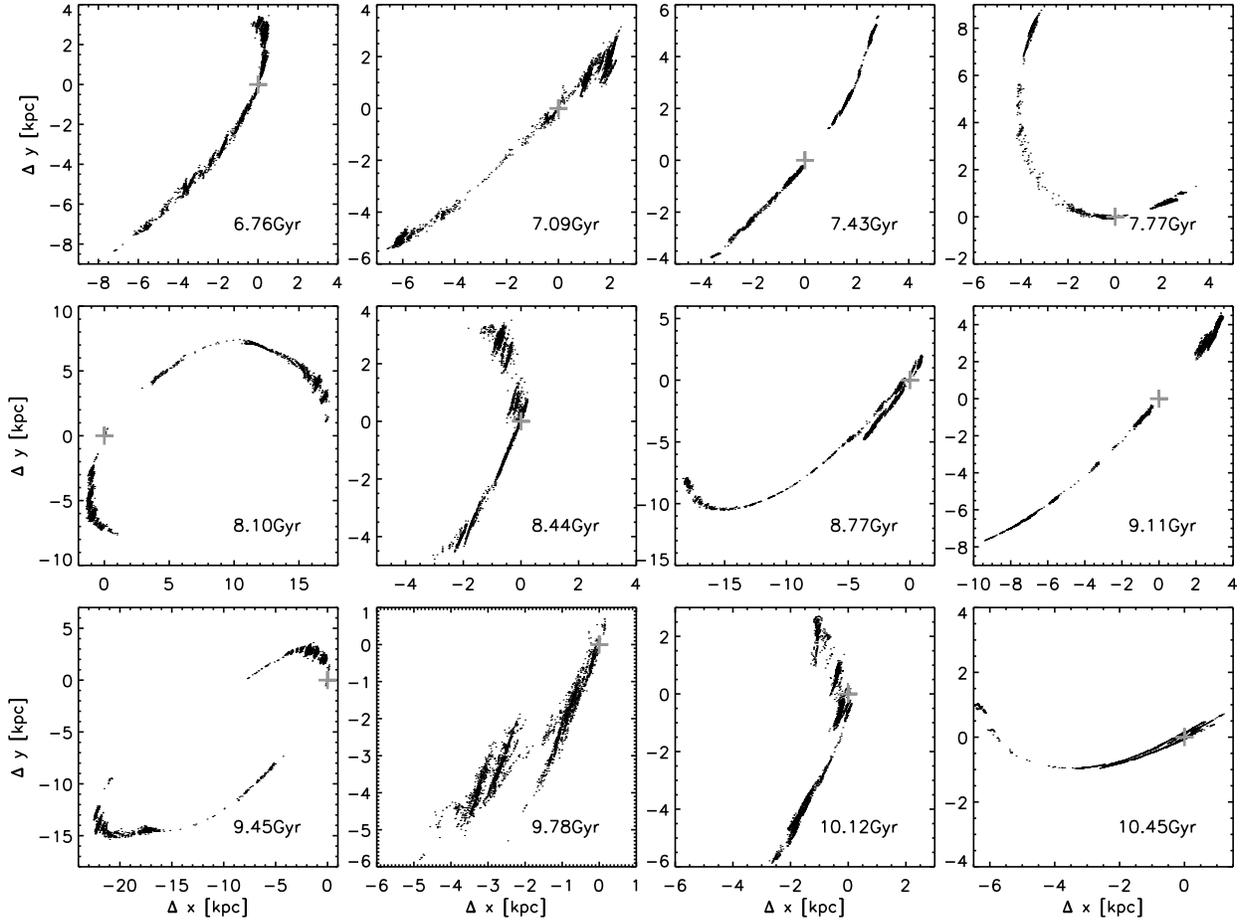} 
   \caption{Final particle positions in the X-Y plane for tidal streams evolved in the presence of the full spectrum of $\Lambda$CDM subhalos. Each
stream begins at differing initial times and positions that were chosen to place all streams at the same ending location in the absence other disturbances  (see the text). 
The age is presented in each panel. Note that each panel 
is centered on the satellite and has a different axis scale.}
 \label{12streams.fig}
 \end{center}
\end{figure*}

Figure~\ref{12streams.fig} summarizes the results of 12 such
simulations with projections of the final particle positions, centered
on the satellite body, onto the x-y plane.  All the simulations were
of the same stream, but started one radial orbital period apart in
order to explore a variety of encounter histories.  The difference
between the panels is most striking.  These simulations were designed
so that all the streams would end with the same shape and orbital
phase (although different lengths) in the absence of interactions with
dark-matter subhalos.  Instead, the streams show a wide variety of
morphologies and orbital phases, that can largely be attributed to
the few encounters with the more massive dark-matter halos (those likely to be
observed as stellar satellites).

Nevertheless, in one way the streams are {\it similar}: they all show
tens of small-scale ($\sim$1kpc and below) gaps, caused by the many
less-massive subhalos --- the missing satellites. The frequency of these
encounters means that these disturbances are apparent in all cases.
In this sense, we agree with \citep{Carlberg2009a} that any observed thin
stream should contain the signatures of missing satellites. However,
it is not clear from Figure ~\ref{12streams.fig} that the mere
existence of old streams rules out the presence of a large population
of low-mass dark-matter subhalos --- many examples of thin, old streams
survive in our simulations.
Indeed, Figure ~\ref{12streams.fig} suggests that
we might expect to find remnants of destroyed streams at surface 
brightness levels below the current level of sensitivity.

The gaps themselves have a striking morphology: their edges are typically not perpendicular to the stream, but rather sit at an angle. This ``slant'' is particularly noticeable for debris at the orbital apocenter. We can trace this appearance in coordinate space to back to the character of the gaps seen  in energy/angular-momentum-space in earlier figures. The edges of the gaps typically have a single orbital energy with a range of angular-momenta. As the debris spreads, the particles sort themselves in energy along the stream and in angular momenta perpendicular to the stream \citep{Johnston2001a}, which results in the angular momenta of particles varying monotonically along the edge of a gap.  
The slant reflects the corresponding monotonic trend in the particles' orbital time periods, which are weakly dependent on angular-momentum.

There are also examples of bizarre morphologies in the streams in
Figure~\ref{12streams.fig}.  
Several streams have discontinuities as
large as several kpc (for example in the middle right-hand and bottom left-hand panel) ---
sufficiently disjointed that theses structure might be detected as separate
streams when we observe them. (Note that the gap between the leading
and trailing streams in each panel is due to the initial conditions in the
simulation and not subhalo interactions.)

\begin{figure}
 \begin{center}
  \vspace{5pt}
   \includegraphics[width=1.05\columnwidth]{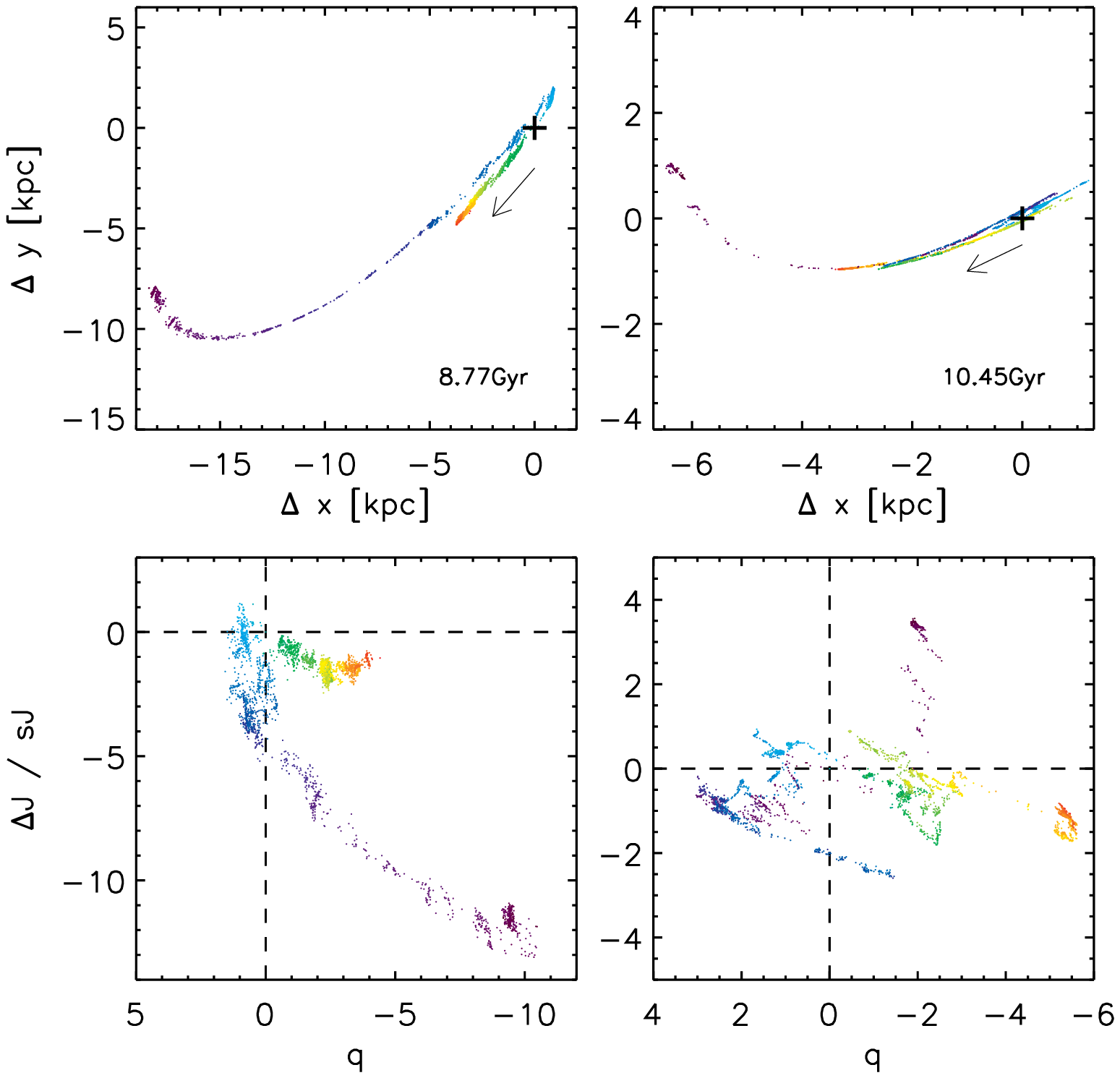} 
   \caption{Two streams showing bifurcations are presented in X-Y and 
   energy/angular-momentum planes. The arrows show the direction of the moving streams.
   The stream particles are color-coded by their initial energy (redder higher, purpler lower).
   The crossing points of dashed lines in the bottom panels show the energy and angular
   momentum of initial central particles (marked as crosses in the top panels).}
 \label{bifur.fig}
 \end{center}
\end{figure}

Some streams even take on a bifurcated appearance, for example in the third panel in the middle row 
and the rightmost panel in the bottom row of Figure~\ref{12streams.fig}. 
The lower panels of Figure \ref{bifur.fig} show the final
energy/angular-momentum distributions for these bifurcated streams, with  
the redder and purpler
particles having initially higher and lower energy respectively.  
Comparing this to the initial distribution in Figure~\ref{scales.fig} shows that in both cases, the energy-angular momentum distributions have been completely flipped by an encounter with a large subhalo, so that the
trailing stream ends up on orbits with shorter time periods and overtakes the leading stream.
A
bifurcation has already been seen in the stellar stream from Sgr
\citep{Belokurov2006b} and could possibly be due to such an encounter.
However, further simulations of an Sgr-scale stream (much longer and
thicker than our Pal-5-like test case) are needed to confirm the
plausibility of this scenario and distinguish it from other
explanations of this bifurcation
\citep[e.g.][]{Fellhauer2006a,Penarrubia2010a}.  We do not pursue this
investigation further here since we are concentrating on encounters of missing satellites with much colder streams.

\section{Discussion }
\label{discussion.sec}

Our results suggest that missing satellites, if present with number
densities predicted by purely dark-matter simulations of structure
formation in a $\Lambda$CDM Universe, should leave their imprint on
cold stellar streams such as Pal 5 and GD-1 in the form of abrupt changes in surface
density and velocity on few degree-scales and below. Moreover, this
signature should be distinct from the few larger-scale deformities expected
to be produced by known satellites. 
In this section, 
we compare our numerical results to current observations and
discuss other possible 
explanations of substructure in stellar streams.

\subsection{The Case of Pal 5: Signatures of Missing Satellites}
\label{sig.sec}

\begin{figure*}
 \begin{center}
  \vspace{5pt}
   \includegraphics[width=2\columnwidth]{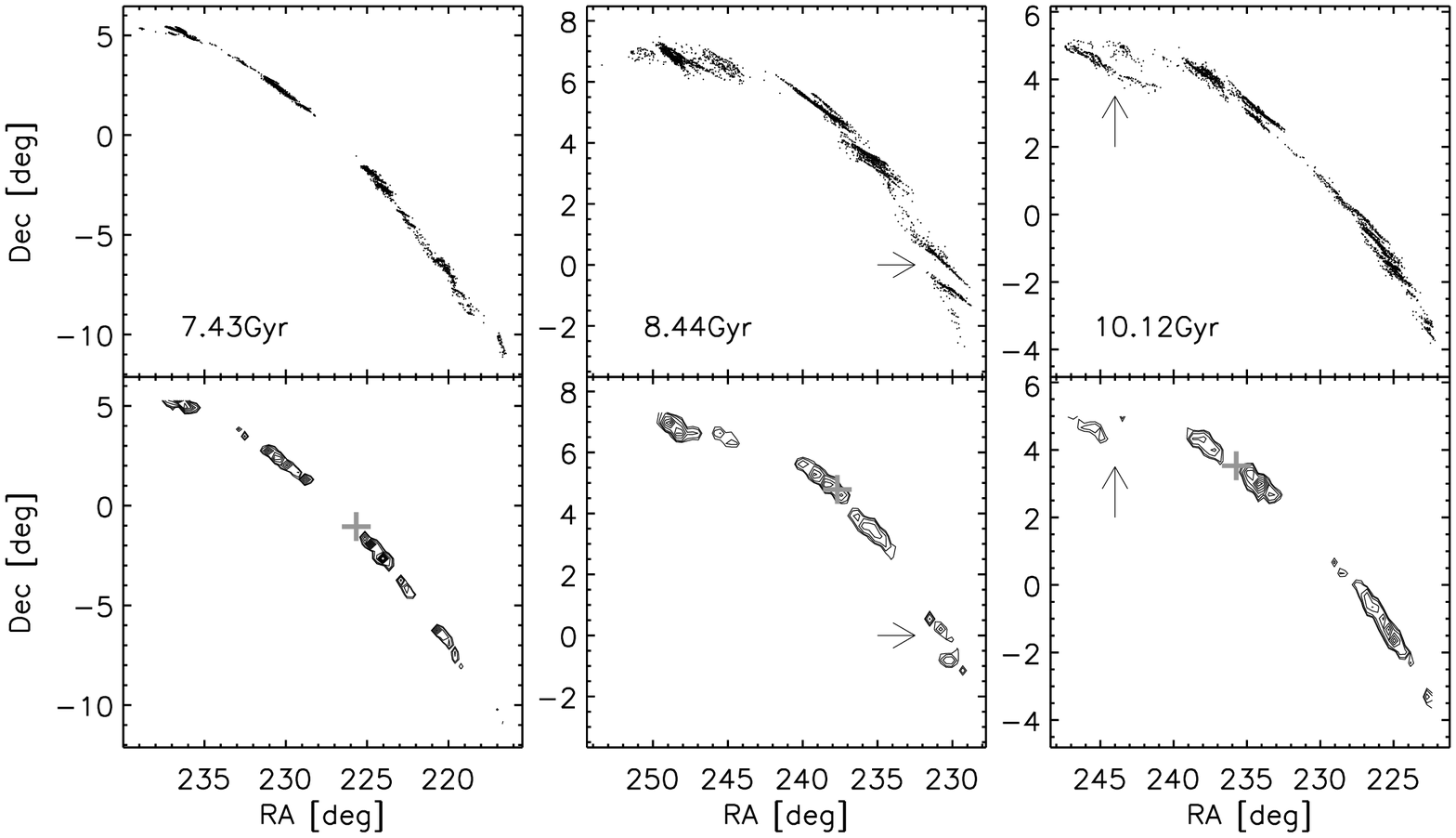} 
   \caption{Particle distribution (upper panels) and surface density maps (lower panels) for  three model streams from Figure~\ref{12streams.fig} projected onto the
   plane of the sky. In the 
   particle plots, many slanted gaps are found which are in some cases still apparent in the contour maps  (indicated by arrows). }
 \label{cntr.fig}
 \end{center}
\end{figure*}

The most straight forward thing to observe about stars in a stream
with current capabilities are their positions on the sky.  In reality,
cold streams like Pal 5 have been discovered as {\it overdensities} of
stars that lie along a restricted sequence in color-magnitude space
(defined by a stellar isochrone), rather than direct identification of
individual stellar members. The identification process
requires some kind of binning or smoothing of the stars in order to decide if a region is
overdense, and this will soften any inhomogeneities in the stream. For
example, Figure 3 of \citet{Odenkirchen2003a,Grillmair2009a} present maps of the region
around Pal 5, in which the star counts have been smoothed with a parabolic and gaussian
kernel.
For comparison, we select three model streams among the ones 
in Figure~\ref{12streams.fig}, count the particles in  $24' \times 24'$ bins 
(chosen to mimc the level of smoothing in the original maps) and
present contours  of surface density in Figure~\ref{cntr.fig}.
\citet{Odenkirchen2003a} found
1,650 stars in a 10$^\circ$ length of stream, 
and we expect comparable surface densities for our particles (of which there are  3,000 in about 25$^\circ$).
Hence we choose the same contour levels  as \citet{Odenkirchen2003a}
at the 1.5, 2, 3, 5-$\sigma$, and higher
levels, with  $\sigma=0.12$ particles/arcmin$^{2}$. 
The largest gaps around RA=227$^\circ$, 243$^\circ$, 
and 231$^\circ$ in each
panel arise from initial conditions (the physical separation of the leading and trailing streams) 
 and not subhalo encounters. Aside from them, we still
find many gaps and clumpy substructures in each stream which --- while somewhat more exaggerated ---
are in rough
agreement with the observed Pal 5 stream.

Moreover, as we discussed in \S~\ref{lcdm.sec}, the gaps in the model 
streams are ``slanted''  and this morphology is also apparent in the contour maps in the middle and right
panels of Figure~\ref{cntr.fig} (arrows in top and bottom panels indicated the relevant gaps). The slants are not as clear in the contour maps as in the particle plots since they are erased out by smoothing, but can become more prominent if the contour levels are varied.  Indeed, in Figure 10 of \citet{Zou2009a}, a similar ``slanted'' gap is seen 
in the bottom-right end of the Pal 5 stream around (RA,Dec)=($227^\circ$,$-2^\circ$). 
The distinctive gap morphology caused by subhalo encounters  could provide a way of distinguishing between competing theories for the origin of stream irregularities (see next section). 
However, a higher density of detected stellar traces is needed to build a convincing case for the existence of slanted gaps.

\begin{figure*}
 \begin{center}
  \vspace{5pt}
   \includegraphics[width=2\columnwidth]{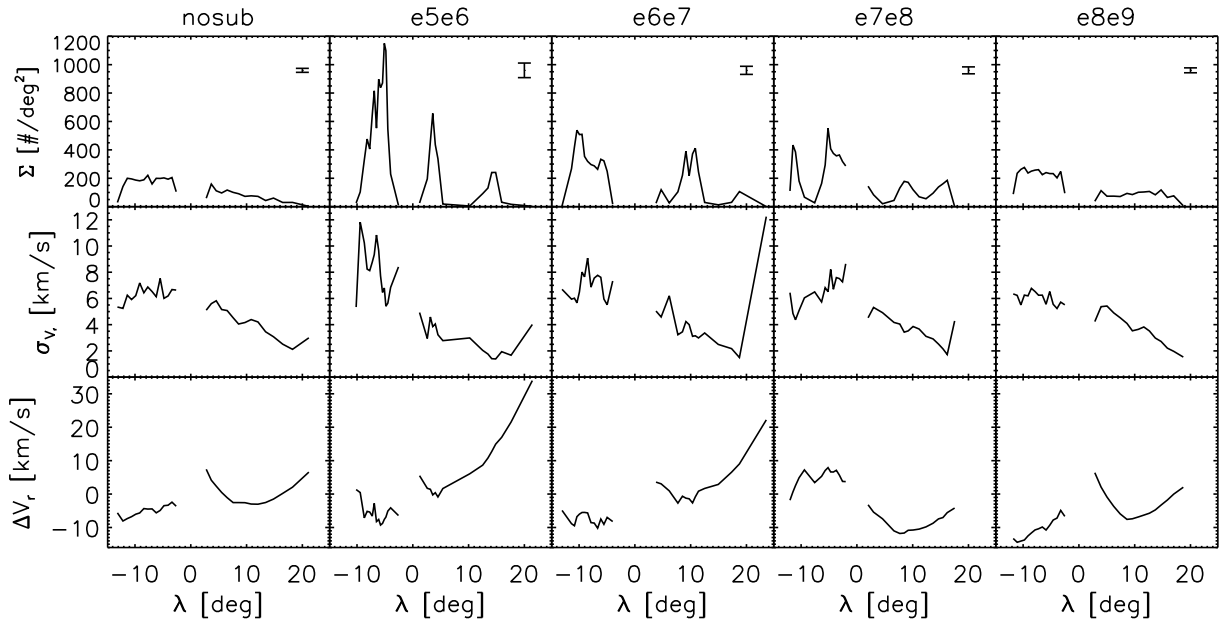} 
   \caption{Top--Bottom: Number density, velocity dispersion, and
     radial velocity profiles along the stream. Left--Right: The stream
     evolved without subhalos, and with subhalos in the mass range $10^5 -
     10^6/10^6 -10^7/10^7 - 10^8/10^8 - 10^9 {\rm \msun}$ are
     presented. $\lambda$ is the angular separation from the host
     satellite center.  Negative and positive $\lambda$ mean trailing and
     leading stream respectively. Mean errors in the counts in the upper panels  
     are indicated by the bar in the
     top-right side of the panels in the top row. }
 \label{profile.fig}
 \end{center}
\end{figure*}

While Figure \ref{cntr.fig} indicates that our simulations which included the full $\Lambda$CDM mass spectrum
appear to be in rough agreement with the level of fluctuations
observed in Pal 5, 
those that contain only the upper end (i.e. corresponding to known Milky Way satellites)
are not so successful.
Figure~\ref{profile.fig} shows the
surface density, radial velocity dispersion, and
radial velocity profiles along the streams illustrated in Figure \ref{varyingmass.fig}, which
represent simulations where the contribution of each mass decade of the
full mass spectrum was isolated.
The profiles are estimated from sets each containing 100 particles
that have been binned in order of increasing angular separation $\lambda$ from the central particle.
The second and third column of 
panels of Figure \ref{profile.fig} --- corresponding to streams disturbed by low-mass subhalos or ``missing satellites'' ---  still reveal surface density fluctuations which are
comparable to the profiles in \citet{Odenkirchen2003a} and \citet{Zou2009a}.
However, fluctuations of the observed
amplitudes and scales are {\it not} seen in our models that included
no satellites (left hand panels), nor in those that included only the ``visible''  satellites
from the top few decades of the $\Lambda$CDM mass spectrum (right-hand panels).

In the future, ideally we would
move beyond broad consistency with the existence of missing
satellites to definitively separating and measuring the contribution
of each decade of the mass spectrum.
This would require sensitivity to a
sufficient number of stars to resolve fluctuations on sub-degree
scales. A comparison of the top panels of Figure \ref{profile.fig}
with those of Figure \ref{varyingmass.fig} illustrates the challenge
of this with our own simulated data. The difference in the nature of
fluctuations caused by subhalos in the range $10^5-10^6 M_\odot$
compared to those caused by the mass decade above is striking in the
particle plots, but the binning of 100 particles to create the profiles smoothes over
$\sim 30^{\circ} {100 \over 3,000} = 1^{\circ}$-scales so that this
signature is less apparent in the density profiles.

Additional information in the form of samples of line-of-sight
velocity measurements of stars in the region of Pal 5's stream are now
being collected, with the first 17 already published by
\citet{Odenkirchen2009a}. Figure \ref{varyingmass.fig} demonstrates that
  the missing satellites scatter stars in velocity space to produce
  local discontinuities of a few km/s on sub-degree scales, while
  Figure \ref{profile.fig} suggests that this scattering may be
  apparent as a small increase in dispersion in averages over a larger
  region. However, finding these signatures observationally remains
challenging as stream stars cannot be definitively identified and any
spectroscopic sample will be contaminated by interlopers.

\subsection{Other Streams}

There are a number of other cold streams with small widths and angular extents (and therefore ages) that rival or even surpass Pal 5 \citep[for example, GD1, NGC5466, Acheron, Cocytos, Lethe, and Styx streams,][]{Grillmair2006a,Grillmair2006c,Fellhauer2007a,Grillmair2009a,Koposov2010a}. Detailed maps of these streams' surface densities -- and of any apparent gaps in them in particular --
could provide complimentary probes of the missing satellites: 
to first order, since these streams have similar widths and all explore the inner few-10 kpc of the Galaxy we expect to find similar numbers of and scales to the signatures of encounters with dark-matter subhalos.

Looking ahead, a more ambitious plan would be to use multiple streams to examine the relative numbers of dark-matter subhalos at different Galactocentric radii.
This interpretation would require rather  better than simple intuitive extrapolations from the current study for the number and scales of gaps.
For example, from the estimates in \citet{Koposov2010a}, while both Pal 5 and GD1 debris sits at similar Galactocentric radii, GD1 orbits slightly farther out (between 18-30kpc from the Galactic Center) and should perhaps therefore encounter fewer missing satellites and contain fewer gaps.
However, although GD1 has a similar width to Pal 5, it has a much longer angular extent ($\sim70^\circ$, or $\sim 40^\circ$ as viewed from the Galactic center): 
naively, a longer extent suggests GD 1 is the older stream, which points in the direction of  {\it more} gaps. 
To further complicate things, GD1 is at pericenter, while Pal 5 is at apocenter, 
and gap scales and morphologies are phase-dependent.
All these differences would have to be taken into account with specific models of each stream before any meaningful comparison could be made.

\subsection{Other Influences on Stream Non-uniformity}
\label{confusion.sec}
There are several other 
 processes that would affect the appearance of clumps
in stellar streams which were not included in our numerical tests.
In some cases, these could plausibly
provide alternative explanations of clump origins which do not require 
the existence of missing satellites.

\begin{description}
\item{Multiple or continuous mass-loss events ---} 
In our simulations, the test particles all started at the same time and orbital phase, mimicking the evolution of
debris form a single mass-loss event and resulting in debris in which the orbital properties of particles varied monotonically with angle along the stream.
As a consequence, subhalos in our simulations directly impacted particles that were not only close physically, but also close in their orbital properties.
In reality, streams are likely to be composed of stars lost over extended periods and during several mass-loss events, so that subhalos would affect debris with a wider range of orbital properties. 
Hence the gaps in energy space created by encounters would not be as clean as those seen in our figures (for example in Figure \ref{varyingmass.fig}). 
This in turn suggests that our simulations have likely overestimated the amplitude (though not physical scales) of fluctuations in the final spatial distribution of the debris:
more realistically, stars from different mass-loss populations would overtake each other across the spatial gaps and soften the contrast.
\item{Asymmetries in the global potential ---} We assumed a purely
  spherical parent potential, neglecting aspherical contributions
  from the disk, bar, spiral arms, and Galactic warp or the (likely)
  triaxiality of the dark-matter halo. \citet{Siegal-Gaskins2008a} have shown
  in particular how the last of these can distort (relatively hot)
  stellar streams from large Galactic satellites. However, we
  anticipate that none of these large asymmetries would leave
  signatures that mimic the degree- and sub-degree scale gaps seen in
  our experiments on colder stellar streams and specifically due to
  direct hits by missing satellites.
\item{Other sources of gravitational shocks ---} Streams can also
  suffer shocks from pericentric encounters with the Galactic
  bar/bulge and when passing through the Galactic disk. 
  Disk shock, which are not taken into account in our models, can 
  destroys globular clusters \citep{Gnedin1999a},
  but are not believed to be the source of discontinuities in Pal 5's stellar streams \citep{Dehnen2004a,Kupper2010a}.
   The
  effect of a disk passage could be enhanced if the stream happened to
  pass directly through a Giant Molecular Cloud (GMC). By
    extrapolating the GMC mass-function, we estimate there to be of
    order one thousand of GMC's within the Solar Circle in the disk
    with masses $< 10^7 \rm \msun $ \citep{Solomon1987a,Williams1997a}, which
    could imply a greater density in this region than dark-matter
    subhalos of comparable scales.  However , although Pal 5 crosses
    the Galactic plane 35 times during the 8.44 Gyrs of our
    simulation, it only passes through the inner disk($< 8$kpc) 8
    times and the closest approach to the Galactic center when it
    crosses the disk is 7.5kpc. Hence, far less encounters of GMCs than 
    the number of direct hits from missing satellites are expected. 
    Overall, we conclude that the
    gravitational shocks from these sources should be much less frequent
    than encounters with subhalos for the Pal 5 stream.
\item{Self gravity of the stream ---} Our test-particle experiments do
  not capture the Jean's instability that could cause clumpy substructure in 
  a stream even in the absence of
  subhalos\citep{Quillen2010a}.  The importance of this effect depends on the density and
  velocity dispersion of the stream as well as the eccentricity of the
  stream orbit.  Simulations including stream self-gravity are needed
  to assess whether it is possible to disentangle the signatures of
  subhalo encounters from Jeans instability.
\item{Epicyclic motions of escaping stars ---} \citet{Kupper2008a,
    Just2009a, Kupper2010a} used analytic estimates and numerical
  experiments to demonstrate that fluctuations in stream density can arise
  from epicyclic motions of escaping stars.  Their experiments showed
  that these fluctuations should be most apparent in cases where the
  stars escape ``evaporatively'' (at low-velocity relative to the
  satellite) and the debris is observed at pericenter. At apocenter of
  their orbit with eccentricity 0.5 (comparable to Pal 5), the
  fluctuations appear to be less than a factor of two in amplitude
and rather regularly spaced, in clear contrast to  the prominent irregularly-spaced
gaps induced in streams by missing satellites in our experiments.
\end{description}

Overall, we anticipate that the mixing of debris from multiple mass loss events could obscure the contrast 
and distinct morphology of the gaps seen in our idealized simulations. However, these softened
fluctuations might actually be {\it more} consistent with those seen in observations of Pal 5's stream.
Of the remaining missing effects, the self-gravity of the stream seems to be the 
dominant factor confusing our interpretation and weakening
our conclusion that the "missing satellites" have been found.

\section{Conclusions}
\label{conclusion.sec}

The goal of this study was to test if a dynamically cold stellar stream
could survive in the presence of dark-matter subhalos and if so, to
characterize observable features in the stream produced by subhalo
encounters.  In particular we were interested in finding signatures of
subhalos that would be considered ``missing satellites'' rather than
those that we already know exist.  We conclude that:
\begin{enumerate}
\item The mere existence of cold stellar streams does not imply the
  absence of missing satellites --- dynamically cold streams can
  survive for many Gyears even when bombarded by subhalos. 
\item Those streams that are observed should contain the imprint of
  past direct impacts from subhalos in the form of gaps in surface
  density and discontinuities in velocities.
\item The frequency and scale of the gaps is dependent on the mass
  spectrum of subhalos and the properties of the stream itself.  In
  the case of Pal 5, there should be observable fluctuations at degree
  and sub-degree scales due to $\sim 90$ encounters with subhalos in the
  mass range $10^5-10^7 M_\odot$ (i.e. the missing satellites), while
  distant encounters with larger subhalos produce less observable
  effects.  Hotter stellar streams, such as Sgr's debris are large
  enough to hide the signatures of the many encounters it suffers with
  missing satellites.
\item Current observations of Pal 5's stream show that its surface
  density profile contains fluctuations on comparable scales to those
  predicted in our simulations, and could be interpreted as 
  direct proof of the existence of missing satellites.
\end{enumerate}

Our study has built on previous works to look at:
effects of individual encounters 
with different mass scales;
the distinct signatures of subhalos in different mass decades; 
the integrated influences 
of subhalo encounters;
and
 comparisons  to
the current observational data.
However, while we have succeeded in outlining the expected scales of
signatures of different decades of the $\Lambda$CDM mass spectrum in
stellar streams, there are still some open questions that need to be
answered before we are confident enough to say that we have definitely
found the missing satellites.  In particular, self-consistent
simulations of Pal 5 disruption including stream self-gravity and
continuous mass loss are needed to clarify to what extent subhalo
signatures could be confused by these effects.  With such simulations
in hand, it would then make sense to develop a multi-dimensional
statistic to compare to all available data in a more robust way. The
promise of current and near-future efforts to map Pal 5 and other
stellar streams are strong motivators for this work.

\acknowledgements
JHY and KVJ were entirely supported by funds from
Columbia University throughout this project.
All simulations were run on Columbia's local computing cluster, {\it Hotfoot}.
DWH was partially supported by the NSF (grant AST-0908357), NASA
(grant NNX08AJ48G), and a research fellowship of the Alexander von
Humboldt Foundation. 
DWH thanks Eric Bell,
Jo Bovy, Andrea Maccio, and Hans-Walter Rix
for useful discussions during the early stages of this project.



\bibliographystyle{apj}
\bibliography{ms}

\begin{thebibliography}{69}
\expandafter\ifx\csname natexlab\endcsname\relax\def\natexlab#1{#1}\fi

\bibitem[{{Belokurov} {et~al.}(2006{\natexlab{a}}){Belokurov}, {Evans},
  {Irwin}, {Hewett}, \& {Wilkinson}}]{Belokurov2006a}
{Belokurov}, V., {Evans}, N.~W., {Irwin}, M.~J., {Hewett}, P.~C., \&
  {Wilkinson}, M.~I. 2006{\natexlab{a}}, \apjl, 637, L29

\bibitem[{{Belokurov} {et~al.}(2006{\natexlab{b}}){Belokurov}, {Zucker},
  {Evans}, {Wilkinson}, {Irwin}, {Hodgkin}, {Bramich}, {Irwin}, {Gilmore},
  {Willman}, {Vidrih}, {Newberg}, {Wyse}, {Fellhauer}, {Hewett}, {Cole},
  {Bell}, {Beers}, {Rockosi}, {Yanny}, {Grebel}, {Schneider}, {Lupton},
  {Barentine}, {Brewington}, {Brinkmann}, {Harvanek}, {Kleinman}, {Krzesinski},
  {Long}, {Nitta}, {Smith}, \& {Snedden}}]{Belokurov2006c}
{Belokurov}, V., {et~al.} 2006{\natexlab{b}}, \apjl, 647, L111

\bibitem[{{Belokurov} {et~al.}(2006{\natexlab{c}}){Belokurov}, {Zucker},
  {Evans}, {Gilmore}, {Vidrih}, {Bramich}, {Newberg}, {Wyse}, {Irwin},
  {Fellhauer}, {Hewett}, {Walton}, {Wilkinson}, {Cole}, {Yanny}, {Rockosi},
  {Beers}, {Bell}, {Brinkmann}, {Ivezi{\'c}}, \& {Lupton}}]{Belokurov2006b}
---. 2006{\natexlab{c}}, \apjl, 642, L137

\bibitem[{{Belokurov} {et~al.}(2007){Belokurov}, {Zucker}, {Evans}, {Kleyna},
  {Koposov}, {Hodgkin}, {Irwin}, {Gilmore}, {Wilkinson}, {Fellhauer},
  {Bramich}, {Hewett}, {Vidrih}, {De Jong}, {Smith}, {Rix}, {Bell}, {Wyse},
  {Newberg}, {Mayeur}, {Yanny}, {Rockosi}, {Gnedin}, {Schneider}, {Beers},
  {Barentine}, {Brewington}, {Brinkmann}, {Harvanek}, {Kleinman}, {Krzesinski},
  {Long}, {Nitta}, \& {Snedden}}]{Belokurov2007a}
---. 2007, \apj, 654, 897

\bibitem[{{Binney}(2008)}]{Binney2008a}
{Binney}, J. 2008, \mnras, 386, L47

\bibitem[{{Binney} \& {Tremaine}(2008)}]{Binney2008b}
{Binney}, J., \& {Tremaine}, S. 2008, {Galactic Dynamics: Second Edition}
  (Princeton University Press)

\bibitem[{{Bovill} \& {Ricotti}(2009)}]{Bovill2009a}
{Bovill}, M.~S., \& {Ricotti}, M. 2009, \apj, 693, 1859

\bibitem[{{Bullock}(2010)}]{Bullock2010a}
{Bullock}, J.~S. 2010, arXiv:1009.4505

\bibitem[{{Bullock} {et~al.}(2010){Bullock}, {Stewart}, {Kaplinghat},
  {Tollerud}, \& {Wolf}}]{Bullock2010b}
{Bullock}, J.~S., {Stewart}, K.~R., {Kaplinghat}, M., {Tollerud}, E.~J., \&
  {Wolf}, J. 2010, \apj, 717, 1043

\bibitem[{{Carlberg}(2009)}]{Carlberg2009a}
{Carlberg}, R.~G. 2009, \apjl, 705, L223

\bibitem[{{Chen} {et~al.}(2003){Chen}, {Kravtsov}, \& {Keeton}}]{Chen2003a}
{Chen}, J., {Kravtsov}, A.~V., \& {Keeton}, C.~R. 2003, \apj, 592, 24

\bibitem[{{Chiba}(2002)}]{Chiba2002a}
{Chiba}, M. 2002, \apj, 565, 17

\bibitem[{{Choi} {et~al.}(2007){Choi}, {Weinberg}, \& {Katz}}]{Choi2007a}
{Choi}, J., {Weinberg}, M.~D., \& {Katz}, N. 2007, \mnras, 381, 987

\bibitem[{{Dehnen} {et~al.}(2004){Dehnen}, {Odenkirchen}, {Grebel}, \&
  {Rix}}]{Dehnen2004a}
{Dehnen}, W., {Odenkirchen}, M., {Grebel}, E.~K., \& {Rix}, H. 2004, \aj, 127,
  2753

\bibitem[{{Diemand} {et~al.}(2007){Diemand}, {Kuhlen}, \&
  {Madau}}]{Diemand2007a}
{Diemand}, J., {Kuhlen}, M., \& {Madau}, P. 2007, \apj, 657, 262

\bibitem[{{Diemand} {et~al.}(2008){Diemand}, {Kuhlen}, {Madau}, {Zemp},
  {Moore}, {Potter}, \& {Stadel}}]{Diemand2008a}
{Diemand}, J., {Kuhlen}, M., {Madau}, P., {Zemp}, M., {Moore}, B., {Potter},
  D., \& {Stadel}, J. 2008, \nat, 454, 735

\bibitem[{{D'Onghia} {et~al.}(2010){D'Onghia}, {Springel}, {Hernquist}, \&
  {Keres}}]{DOnghia2010a}
{D'Onghia}, E., {Springel}, V., {Hernquist}, L., \& {Keres}, D. 2010, \apj,
  709, 1138

\bibitem[{{Eyre}(2010)}]{Eyre2010a}
{Eyre}, A. 2010, \mnras, 403, 1999

\bibitem[{{Fellhauer} {et~al.}(2007){Fellhauer}, {Evans}, {Belokurov},
  {Wilkinson}, \& {Gilmore}}]{Fellhauer2007a}
{Fellhauer}, M., {Evans}, N.~W., {Belokurov}, V., {Wilkinson}, M.~I., \&
  {Gilmore}, G. 2007, \mnras, 380, 749

\bibitem[{{Fellhauer} {et~al.}(2006){Fellhauer}, {Belokurov}, {Evans},
  {Wilkinson}, {Zucker}, {Gilmore}, {Irwin}, {Bramich}, {Vidrih}, {Wyse},
  {Beers}, \& {Brinkmann}}]{Fellhauer2006a}
{Fellhauer}, M., {et~al.} 2006, \apj, 651, 167

\bibitem[{{Gnedin} {et~al.}(1999){Gnedin}, {Lee}, \& {Ostriker}}]{Gnedin1999a}
{Gnedin}, O.~Y., {Lee}, H.~M., \& {Ostriker}, J.~P. 1999, \apj, 522, 935

\bibitem[{{Goerdt} {et~al.}(2007){Goerdt}, {Gnedin}, {Moore}, {Diemand}, \&
  {Stadel}}]{Goerdt2007a}
{Goerdt}, T., {Gnedin}, O.~Y., {Moore}, B., {Diemand}, J., \& {Stadel}, J.
  2007, \mnras, 375, 191

\bibitem[{{Grillmair}(2006)}]{Grillmair2006d}
{Grillmair}, C.~J. 2006, \apjl, 645, L37

\bibitem[{{Grillmair}(2009)}]{Grillmair2009a}
---. 2009, \apj, 693, 1118

\bibitem[{{Grillmair} \& {Dionatos}(2006{\natexlab{a}})}]{Grillmair2006b}
{Grillmair}, C.~J., \& {Dionatos}, O. 2006{\natexlab{a}}, \apjl, 641, L37

\bibitem[{{Grillmair} \& {Dionatos}(2006{\natexlab{b}})}]{Grillmair2006c}
---. 2006{\natexlab{b}}, \apjl, 643, L17

\bibitem[{{Grillmair} \& {Johnson}(2006)}]{Grillmair2006a}
{Grillmair}, C.~J., \& {Johnson}, R. 2006, \apjl, 639, L17

\bibitem[{{Helmi} \& {White}(1999)}]{Helmi1999a}
{Helmi}, A., \& {White}, S.~D.~M. 1999, \mnras, 307, 495

\bibitem[{{Hernquist} \& {Ostriker}(1992)}]{Hernquist1992a}
{Hernquist}, L., \& {Ostriker}, J.~P. 1992, \apj, 386, 375

\bibitem[{{Hooper} {et~al.}(2007){Hooper}, {Kaplinghat}, {Strigari}, \&
  {Zurek}}]{Hooper2007a}
{Hooper}, D., {Kaplinghat}, M., {Strigari}, L.~E., \& {Zurek}, K.~M. 2007,
  \prd, 76, 103515

\bibitem[{{Ibata} {et~al.}(2002){Ibata}, {Lewis}, {Irwin}, \&
  {Quinn}}]{Ibata2002a}
{Ibata}, R.~A., {Lewis}, G.~F., {Irwin}, M.~J., \& {Quinn}, T. 2002, \mnras,
  332, 915

\bibitem[{{Irwin} {et~al.}(2007){Irwin}, {Belokurov}, {Evans}, {Ryan-Weber},
  {de Jong}, {Koposov}, {Zucker}, {Hodgkin}, {Gilmore}, {Prema}, {Hebb},
  {Begum}, {Fellhauer}, {Hewett}, {Kennicutt}, {Wilkinson}, {Bramich},
  {Vidrih}, {Rix}, {Beers}, {Barentine}, {Brewington}, {Harvanek},
  {Krzesinski}, {Long}, {Nitta}, \& {Snedden}}]{Irwin2007a}
{Irwin}, M.~J., {et~al.} 2007, \apjl, 656, L13

\bibitem[{{Johnston}(1998)}]{Johnston1998a}
{Johnston}, K.~V. 1998, \apj, 495, 297

\bibitem[{{Johnston} {et~al.}(2005){Johnston}, {Law}, \&
  {Majewski}}]{Johnston2005a}
{Johnston}, K.~V., {Law}, D.~R., \& {Majewski}, S.~R. 2005, \apj, 619, 800

\bibitem[{{Johnston} {et~al.}(2001){Johnston}, {Sackett}, \&
  {Bullock}}]{Johnston2001a}
{Johnston}, K.~V., {Sackett}, P.~D., \& {Bullock}, J.~S. 2001, \apj, 557, 137

\bibitem[{{Johnston} {et~al.}(2002){Johnston}, {Spergel}, \&
  {Haydn}}]{Johnston2002a}
{Johnston}, K.~V., {Spergel}, D.~N., \& {Haydn}, C. 2002, \apj, 570, 656

\bibitem[{{Johnston} {et~al.}(1999){Johnston}, {Zhao}, {Spergel}, \&
  {Hernquist}}]{Johnston1999a}
{Johnston}, K.~V., {Zhao}, H., {Spergel}, D.~N., \& {Hernquist}, L. 1999,
  \apjl, 512, L109

\bibitem[{{Just} {et~al.}(2009){Just}, {Berczik}, {Petrov}, \&
  {Ernst}}]{Just2009a}
{Just}, A., {Berczik}, P., {Petrov}, M.~I., \& {Ernst}, A. 2009, \mnras, 392,
  969

\bibitem[{{Keeton} \& {Moustakas}(2009)}]{Keeton2009a}
{Keeton}, C.~R., \& {Moustakas}, L.~A. 2009, \apj, 699, 1720

\bibitem[{{Klypin} {et~al.}(1999){Klypin}, {Kravtsov}, {Valenzuela}, \&
  {Prada}}]{Klypin1999a}
{Klypin}, A., {Kravtsov}, A.~V., {Valenzuela}, O., \& {Prada}, F. 1999, \apj,
  522, 82

\bibitem[{{Koposov} {et~al.}(2007){Koposov}, {de Jong}, {Belokurov}, {Rix},
  {Zucker}, {Evans}, {Gilmore}, {Irwin}, \& {Bell}}]{Koposov2007a}
{Koposov}, S., {et~al.} 2007, \apj, 669, 337

\bibitem[{{Koposov} {et~al.}(2008){Koposov}, {Belokurov}, {Evans}, {Hewett},
  {Irwin}, {Gilmore}, {Zucker}, {Rix}, {Fellhauer}, {Bell}, \&
  {Glushkova}}]{Koposov2008a}
---. 2008, \apj, 686, 279

\bibitem[{{Koposov} {et~al.}(2010){Koposov}, {Rix}, \& {Hogg}}]{Koposov2010a}
{Koposov}, S.~E., {Rix}, H., \& {Hogg}, D.~W. 2010, \apj, 712, 260

\bibitem[{{K{\"u}pper} {et~al.}(2010){K{\"u}pper}, {Kroupa}, {Baumgardt}, \&
  {Heggie}}]{Kupper2010a}
{K{\"u}pper}, A.~H.~W., {Kroupa}, P., {Baumgardt}, H., \& {Heggie}, D.~C. 2010,
  \mnras, 401, 105

\bibitem[{{K{\"u}pper} {et~al.}(2008){K{\"u}pper}, {MacLeod}, \&
  {Heggie}}]{Kupper2008a}
{K{\"u}pper}, A.~H.~W., {MacLeod}, A., \& {Heggie}, D.~C. 2008, \mnras, 387,
  1248

\bibitem[{{Lauchner} {et~al.}(2006){Lauchner}, {Powell}, \&
  {Wilhelm}}]{Lauchner2006a}
{Lauchner}, A., {Powell}, Jr., W.~L., \& {Wilhelm}, R. 2006, \apjl, 651, L33

\bibitem[{{Law} \& {Majewski}(2010)}]{Law2010a}
{Law}, D.~R., \& {Majewski}, S.~R. 2010, \apj, 714, 229

\bibitem[{{Metcalf} {et~al.}(2004){Metcalf}, {Moustakas}, {Bunker}, \&
  {Parry}}]{Metcalf2004a}
{Metcalf}, R.~B., {Moustakas}, L.~A., {Bunker}, A.~J., \& {Parry}, I.~R. 2004,
  \apj, 607, 43

\bibitem[{{Metcalf} \& {Zhao}(2002)}]{Metcalf2002a}
{Metcalf}, R.~B., \& {Zhao}, H. 2002, \apjl, 567, L5

\bibitem[{{Moore} {et~al.}(1999){Moore}, {Ghigna}, {Governato}, {Lake},
  {Quinn}, {Stadel}, \& {Tozzi}}]{Moore1999a}
{Moore}, B., {Ghigna}, S., {Governato}, F., {Lake}, G., {Quinn}, T., {Stadel},
  J., \& {Tozzi}, P. 1999, \apjl, 524, L19

\bibitem[{{Moustakas} \& {Metcalf}(2003)}]{Moustakas2003a}
{Moustakas}, L.~A., \& {Metcalf}, R.~B. 2003, \mnras, 339, 607

\bibitem[{{Navarro} {et~al.}(1996){Navarro}, {Frenk}, \&
  {White}}]{Navarro1996a}
{Navarro}, J.~F., {Frenk}, C.~S., \& {White}, S.~D.~M. 1996, \apj, 462, 563

\bibitem[{{Odenkirchen} {et~al.}(2009){Odenkirchen}, {Grebel}, {Kayser}, {Rix},
  \& {Dehnen}}]{Odenkirchen2009a}
{Odenkirchen}, M., {Grebel}, E.~K., {Kayser}, A., {Rix}, H., \& {Dehnen}, W.
  2009, \aj, 137, 3378

\bibitem[{{Odenkirchen} {et~al.}(2001){Odenkirchen}, {Grebel}, {Rockosi},
  {Dehnen}, {Ibata}, {Rix}, {Stolte}, {Wolf}, {Anderson}, {Bahcall},
  {Brinkmann}, {Csabai}, {Hennessy}, {Hindsley}, {Ivezi{\'c}}, {Lupton},
  {Munn}, {Pier}, {Stoughton}, \& {York}}]{Odenkirchen2001a}
{Odenkirchen}, M., {et~al.} 2001, \apjl, 548, L165

\bibitem[{{Odenkirchen} {et~al.}(2003){Odenkirchen}, {Grebel}, {Dehnen}, {Rix},
  {Yanny}, {Newberg}, {Rockosi}, {Mart{\'{\i}}nez-Delgado}, {Brinkmann}, \&
  {Pier}}]{Odenkirchen2003a}
---. 2003, \aj, 126, 2385

\bibitem[{{Pe{\~n}arrubia} {et~al.}(2010){Pe{\~n}arrubia}, {Belokurov},
  {Evans}, {Mart{\'{\i}}nez-Delgado}, {Gilmore}, {Irwin}, {Niederste-Ostholt},
  \& {Zucker}}]{Penarrubia2010a}
{Pe{\~n}arrubia}, J., {Belokurov}, V., {Evans}, N.~W.,
  {Mart{\'{\i}}nez-Delgado}, D., {Gilmore}, G., {Irwin}, M.,
  {Niederste-Ostholt}, M., \& {Zucker}, D.~B. 2010, \mnras, 408, L26

\bibitem[{{Quillen} \& {Comparetta}(2010)}]{Quillen2010a}
{Quillen}, A.~C., \& {Comparetta}, J. 2010, arXiv:1002.4870

\bibitem[{{Quinn} {et~al.}(2008){Quinn}, {Wilkinson}, \& {Irwin}}]{Quinn2008a}
{Quinn}, D.~P., {Wilkinson}, M.~I., \& {Irwin}, M.~J. 2008, Astronomische
  Nachrichten, 329, 1071

\bibitem[{{Riehm} {et~al.}(2009){Riehm}, {Zackrisson}, {M{\"o}rtsell}, \&
  {Wiik}}]{Riehm2009a}
{Riehm}, T., {Zackrisson}, E., {M{\"o}rtsell}, E., \& {Wiik}, K. 2009, \apj,
  700, 1552

\bibitem[{{Siegal-Gaskins} \& {Valluri}(2008)}]{Siegal-Gaskins2008a}
{Siegal-Gaskins}, J.~M., \& {Valluri}, M. 2008, \apj, 681, 40

\bibitem[{{Solomon} {et~al.}(1987){Solomon}, {Rivolo}, {Barrett}, \&
  {Yahil}}]{Solomon1987a}
{Solomon}, P.~M., {Rivolo}, A.~R., {Barrett}, J., \& {Yahil}, A. 1987, \apj,
  319, 730

\bibitem[{{Springel} {et~al.}(2008){Springel}, {Wang}, {Vogelsberger},
  {Ludlow}, {Jenkins}, {Helmi}, {Navarro}, {Frenk}, \& {White}}]{Springel2008a}
{Springel}, V., {et~al.} 2008, \mnras, 391, 1685

\bibitem[{{Tollerud} {et~al.}(2008){Tollerud}, {Bullock}, {Strigari}, \&
  {Willman}}]{Tollerud2008a}
{Tollerud}, E.~J., {Bullock}, J.~S., {Strigari}, L.~E., \& {Willman}, B. 2008,
  \apj, 688, 277

\bibitem[{{Walsh} {et~al.}(2007){Walsh}, {Jerjen}, \& {Willman}}]{Walsh2007a}
{Walsh}, S.~M., {Jerjen}, H., \& {Willman}, B. 2007, \apjl, 662, L83

\bibitem[{{Williams} \& {McKee}(1997)}]{Williams1997a}
{Williams}, J.~P., \& {McKee}, C.~F. 1997, \apj, 476, 166

\bibitem[{{Willman} {et~al.}(2005){Willman}, {Dalcanton}, {Martinez-Delgado},
  {West}, {Blanton}, {Hogg}, {Barentine}, {Brewington}, {Harvanek}, {Kleinman},
  {Krzesinski}, {Long}, {Neilsen}, {Nitta}, \& {Snedden}}]{Willman2005a}
{Willman}, B., {et~al.} 2005, \apjl, 626, L85

\bibitem[{{Xu} {et~al.}(2009){Xu}, {Mao}, {Wang}, {Springel}, {Gao}, {White},
  {Frenk}, {Jenkins}, {Li}, \& {Navarro}}]{Xu2009a}
{Xu}, D.~D., {et~al.} 2009, \mnras, 398, 1235

\bibitem[{{Zou} {et~al.}(2009){Zou}, {Wu}, {Ma}, \& {Zhou}}]{Zou2009a}
{Zou}, H., {Wu}, Z., {Ma}, J., \& {Zhou}, X. 2009, Research in Astronomy and
  Astrophysics, 9, 1131

\bibitem[{{Zucker} {et~al.}(2006){Zucker}, {Belokurov}, {Evans}, {Wilkinson},
  {Irwin}, {Sivarani}, {Hodgkin}, {Bramich}, {Irwin}, {Gilmore}, {Willman},
  {Vidrih}, {Fellhauer}, {Hewett}, {Beers}, {Bell}, {Grebel}, {Schneider},
  {Newberg}, {Wyse}, {Rockosi}, {Yanny}, {Lupton}, {Smith}, {Barentine},
  {Brewington}, {Brinkmann}, {Harvanek}, {Kleinman}, {Krzesinski}, {Long},
  {Nitta}, \& {Snedden}}]{Zucker2006a}
{Zucker}, D.~B., {et~al.} 2006, \apjl, 643, L103

\end{thebibliography}



\end{document}